\documentclass[aps, prd, superscriptaddress,nofootinbib, preprint]{revtex4-1}

\usepackage{amsmath,amssymb,amsfonts,bm,slashed,graphicx,array,multirow,xspace,makecell}
\usepackage{placeins}
\usepackage{setspace}
\usepackage{enumitem}
\usepackage{hhline}
\usepackage[labelfont=bf,font={small,stretch=1}, justification=centerlast]{caption}
\usepackage{subcaption}
\usepackage{tikz}
\usepackage[official]{eurosym}
\usepackage{listings}
\usepackage{breakurl}
\usepackage{comment}
\definecolor{red}{rgb}{0.9, 0,0}
\definecolor{cerulean}{rgb}{0., 0.42,0.9}
\definecolor{navy}{rgb}{0.05, 0.05,0.8}

\definecolor{nicered}{rgb}{0.7,0.1,0.1}
\definecolor{nicegreen}{rgb}{0.1,0.5,0.1}
\definecolor{niceblue}{rgb}{0.1,0.1,0.7}
\usepackage[bookmarks=false]{hyperref}
\hypersetup{colorlinks,citecolor= nicegreen,linkcolor= niceblue}

\newcommand{\CODEXb}{\mbox{CODEX-b}\xspace}
\newcommand{\CODEXbeta}{\mbox{CODEX-$\beta$}\xspace}

\newcommand{\nn}{{\nonumber}}

\newcommand{\sigmaopt}[1]{\sigma_\text{opt}^{(#1)}}
\newcommand{\fbar}{\bar{f}}
\newcommand{\dsigma}{\delta\sigma}

\usepackage{extramarks}

\allowdisplaybreaks

\makeatletter
\g@addto@macro\bfseries{\boldmath}
\makeatother

\let\origfootnote\footnote
\renewcommand{\footnote}[1]{%
   \begingroup
   \renewcommand{\baselinestretch}{1}%
   \origfootnote{#1}%
   \endgroup}
\begin{document}

\newcommand\cpp[1]{{\lstinline!#1!}}  
\newcommand\cpppragma[1]{{\CPPcommentstyle#1}}
\newcommand\yaml[1]{{\lstset{style=yaml}\lstinline!#1!\lstset{style=cpp}}}
\newcommand\yamlvalue[1]{{\YAMLvaluestyle\ttfamily#1}}
\newcommand\term[1]{{\lstset{style=terminal}\lstinline!#1!\lstset{style=cpp}}}
\newcommand\fortran[1]{{\lstset{style=fortran}\lstinline!#1!\lstset{style=cpp}}}
\newcommand\py[1]{{\lstset{style=python}\lstinline!#1!\lstset{style=cpp}}}
\newcommand\customtilde{{\raisebox{0.2ex}{\scalebox{0.6}{\boldmath$\sim$}}}}
\newcommand\mathematica[1]{{\lstset{style=Mathematica}\lstinline!#1!\lstset{style=cpp}}}

\lstnewenvironment{lstlistingyaml}{\lstset{style=yaml}}{\lstset{style=cpp}}
\lstnewenvironment{lstlistingterm}{\lstset{style=terminal}}{\lstset{style=cpp}}
\lstnewenvironment{lstlistingfortran}{\lstset{style=fortran}}{\lstset{style=cpp}}
\lstnewenvironment{lstcpp}{\lstset{style=cpp}}{\lstset{style=cpp}}
\lstnewenvironment{lstcppalt}{\lstset{style=cppalt}}{\lstset{style=cpp}}
\lstnewenvironment{lstcppnum}{\lstset{style=cppnum}}{\lstset{style=cpp}}
\lstnewenvironment{lstyaml}{\lstset{style=yaml}}{\lstset{style=cpp}}
\lstnewenvironment{lstterm}{\lstset{style=terminal}}{\lstset{style=cpp}}
\lstnewenvironment{lsttermalt}{\lstset{style=terminalalt}}{\lstset{style=cpp}}
\lstnewenvironment{lsttext}{\lstset{style=text}}{\lstset{style=cpp}}
\lstnewenvironment{lstfortran}{\lstset{style=fortran}}{\lstset{style=cpp}}
\lstnewenvironment{lstpy}{\lstset{style=python}}{\lstset{style=cpp}}
\lstnewenvironment{lstmathematica}{\lstset{style=mathematica}}{\lstset{style=cpp}}

\newcommand{\tmpname}{}
\newcommand{\tmplistingname}{}
\makeatletter
\newif\ifATOlabelname
\lst@Key{labelname}{Listing}{\def\ATOlabelname{#1}\global\ATOlabelnametrue}
\makeatother
\lstnewenvironment{lstcpplabel}[1][]{
  \lstset{style=cpp,#1} 
  \ifATOlabelname
    \renewcommand{\tmpname}{\lstlistingname}
    \renewcommand{\tmplistingname}{\lstlistlistingname}
    \renewcommand{\lstlistingname}{\ATOlabelname}
    \renewcommand{\lstlistlistingname}{List of \lstlistingname s}
  \fi
}{
  \renewcommand{\lstlistingname}{\tmpname}
  \renewcommand{\lstlistlistingname}{\tmplistingname}
  \lstset{style=cpp}
}
\definecolor{solarized@base03}{HTML}{002B36}
\definecolor{solarized@base02}{HTML}{073642}
\definecolor{solarized@base01}{HTML}{586e75}
\definecolor{solarized@base00}{HTML}{657b83}
\definecolor{solarized@base0}{HTML}{839496}
\definecolor{solarized@base1}{HTML}{93a1a1}
\definecolor{solarized@base2}{HTML}{EEE8D5}
\definecolor{solarized@base3}{HTML}{FDF6E3}
\definecolor{solarized@yellow}{HTML}{B58900}
\definecolor{solarized@orange}{HTML}{CB4B16}
\definecolor{solarized@red}{HTML}{DC322F}
\definecolor{solarized@magenta}{HTML}{D33682}
\definecolor{solarized@violet}{HTML}{6C71C4}
\definecolor{solarized@blue}{HTML}{268BD2}
\definecolor{solarized@cyan}{HTML}{2AA198}
\definecolor{solarized@green}{HTML}{859900}
\definecolor{darkred}{HTML}{550003}
\definecolor{darkgreen}{HTML}{00AA00}
\newcommand\YAMLcolonstyle{\footnotesize\color{solarized@red}\mdseries}
\newcommand\YAMLstringstyle{\footnotesize\color{solarized@green}\mdseries}
\newcommand\YAMLkeystyle{\footnotesize\color{solarized@blue}\ttfamily}
\newcommand\YAMLvaluestyle{\footnotesize\color{blue}\mdseries}
\newcommand\ProcessThreeDashes{\llap{\color{cyan}\mdseries-{-}-}}
\newcommand\CPPplainstyle{\footnotesize\ttfamily}
\newcommand\CPPkeywordstyle{\color{solarized@orange}\footnotesize\ttfamily}
\newcommand\CPPidentifierstyle{\color{solarized@blue}\footnotesize\ttfamily}
\newcommand\CPPcommentstyle{\color{solarized@violet}\footnotesize\ttfamily}
\newcommand\CPPdirectivestyle{\color{solarized@magenta}\footnotesize\ttfamily}
\newcommand\termplainstyle{\footnotesize\ttfamily}

\newcommand\processLongMacroDelimiter
{%
\CPPdirectivestyle%
\#define%
}

\newcommand\processCPPCOMMENT
{%
\CPPcommentstyle%
{//}%
}

\newcommand\processCPPTRIPCOMMENT
{%
\CPPcommentstyle%
{///}%
}

\lstdefinestyle{cpp}
{
  language=C++,
  basicstyle=\footnotesize\ttfamily,
  basewidth={0.53em,0.44em}, 
  numbers=none,
  tabsize=2,
  breaklines=true,
  escapeinside={@}{@},
  showstringspaces=false,
  numberstyle=\tiny\color{solarized@base01},
  keywordstyle=\color{solarized@orange},
  stringstyle=\color{solarized@red}\ttfamily,
  identifierstyle=\color{solarized@blue},
  commentstyle=\CPPcommentstyle,
  directivestyle=\CPPdirectivestyle,
  emphstyle=\color{solarized@green},
  frame=single,
  rulecolor=\color{solarized@base2},
  rulesepcolor=\color{solarized@base2},
  literate={~} {\customtilde}1,
  moredelim=*[directive]\ \ \#,
  moredelim=*[directive]\ \ \ \ \#,
  }

\lstdefinestyle{cppalt}
{
  language=C++,
  basicstyle=\footnotesize\ttfamily,
  basewidth={0.53em,0.44em}, 
  numbers=none,
  tabsize=2,
  breaklines=true,
  escapeinside={*@}{@*},
  showstringspaces=false,
  numberstyle=\tiny\color{solarized@base01},
  keywordstyle=\color{solarized@orange},
  stringstyle=\color{solarized@red}\ttfamily,
  identifierstyle=\color{solarized@blue},
  commentstyle=\CPPcommentstyle,
  directivestyle=\CPPdirectivestyle,
  emphstyle=\color{solarized@green},
  frame=single,
  rulecolor=\color{solarized@base2},
  rulesepcolor=\color{solarized@base2},
  literate={~}{\customtilde}1,
  moredelim=**[is][\processLongMacroDelimiter]{BeginLongMacro}{EndLongMacro} 
}

\lstdefinestyle{cppnum}
{
  language=C++,
  basicstyle=\footnotesize\ttfamily,
  basewidth={0.53em,0.44em}, 
  numbers=none,
  tabsize=2,
  breaklines=true,
  escapeinside={@}{@},
  numberstyle=\tiny\color{solarized@base01},
  showstringspaces=false,
  numberstyle=\tiny\color{solarized@base01},
  keywordstyle=\color{solarized@orange},
  stringstyle=\color{solarized@red}\ttfamily,
  identifierstyle=\color{solarized@blue},
  commentstyle=\CPPcommentstyle,
  directivestyle=\CPPdirectivestyle,
  emphstyle=\color{solarized@green},
  frame=single,
  rulecolor=\color{solarized@base2},
  rulesepcolor=\color{solarized@base2},
  literate={~} {\customtilde}1,
  moredelim=*[directive]\ \ \#,
  moredelim=*[directive]\ \ \ \ \#
}

\lstdefinestyle{python}
{
  language=Python,
  basicstyle=\linespread{1.0}\footnotesize\ttfamily,
  basewidth={0.53em,0.44em},
  numbers=none,
  tabsize=2,
  breaklines=true,
  escapeinside={@}{@},
  showstringspaces=false,
  numberstyle=\tiny\color{solarized@base01},
  keywordstyle=\color{blue},
  stringstyle=\color{orange}\ttfamily,
  identifierstyle=\color{darkred},
  commentstyle=\color{purple},
  emphstyle=\color{green},
  frame=single,
  rulecolor=\color{solarized@base2},
  rulesepcolor=\color{solarized@base2},
  literate = {~}{\customtilde}1
             {\ as\ }{{\color{blue}\ as\ \color{black}}}3
}

\lstdefinestyle{fortran}
{
  language=Fortran,
  basicstyle=\footnotesize\ttfamily,
  basewidth={0.53em,0.44em},
  numbers=none,
  tabsize=2,
  breaklines=true,
  escapeinside={@}{@},
  showstringspaces=false,
  numberstyle=\tiny\color{solarized@base01},
  keywordstyle=\color{blue},
  stringstyle=\color{orange}\ttfamily,
  identifierstyle=\color{Periwinkle},
  commentstyle=\color{purple},
  emphstyle=\color{green},
  morekeywords={and, or, true, false},
  frame=single,
  rulecolor=\color{solarized@base2},
  rulesepcolor=\color{solarized@base2},
  literate={~}{\customtilde}1
}

\lstdefinestyle{terminal}
{
  language=bash,
  basicstyle=\termplainstyle,
  numbers=none,
  tabsize=2,
  breaklines=true,
  escapeinside={@}{@},
  frame=single,
  showstringspaces=false,
  numberstyle=\tiny\color{solarized@base01},
  keywordstyle=\color{solarized@orange},
  stringstyle=\color{solarized@red}\ttfamily,
  identifierstyle=\color{black},
  commentstyle=\color{solarized@violet},
  emphstyle=\color{solarized@green},
  frame=single,
  rulecolor=\color{solarized@base2},
  rulesepcolor=\color{solarized@base2},
  morekeywords={gambit, cmake, make, mkdir},
  deletekeywords={test},
  literate = {\ gambit}{{\ }{\color{black}}gambit}7
             {/gambit}{{/}{\color{black}}gambit}6
             {gambit/}{{\color{black}}gambit{/}}6
             {/include}{{/}{\color{black}}include}8
             {cmake/}{{\color{black}}cmake/}6
             {.cmake}{{.}{\color{black}}cmake}6
             {~}{\customtilde}1
}

\lstdefinestyle{terminalalt}
{
  language=bash,
  basicstyle=\footnotesize\ttfamily,
  numbers=none,
  tabsize=2,
  breaklines=true,
  escapeinside={*@}{@*},
  frame=single,
  showstringspaces=false,
  numberstyle=\tiny\color{solarized@base01},
  keywordstyle=\color{solarized@orange},
  stringstyle=\color{solarized@red}\ttfamily,
  identifierstyle=\color{black},
  commentstyle=\color{solarized@violet},
  emphstyle=\color{solarized@green},
  frame=single,
  rulecolor=\color{solarized@base2},
  rulesepcolor=\color{solarized@base2},
  morekeywords={gambit, cmake, make, mkdir},
  deletekeywords={test},
  literate = {\ gambit}{{\ }{\color{black}}gambit}7
             {/gambit}{{/}{\color{black}}gambit}6
             {gambit/}{{\color{black}}gambit{/}}6
             {/include}{{/}{\color{black}}include}8
             {cmake/}{{\color{black}}cmake/}6
             {.cmake}{{.}{\color{black}}cmake}6
             {~}{\customtilde}1
}

\lstdefinestyle{text}
{
  language={},
  basicstyle=\footnotesize\ttfamily,
  identifierstyle=\color{black},
  numbers=none,
  tabsize=2,
  breaklines=true,
  escapeinside={*@}{@*},
  showstringspaces=false,
  frame=single,
  rulecolor=\color{solarized@base2},
  rulesepcolor=\color{solarized@base2},
  literate={~}{\customtilde}1
}

\lstdefinestyle{yaml}
{
  language=bash,
  escapeinside={@}{@},
  keywords={true,false,null},
  otherkeywords={},
  keywordstyle=\color{solarized@base0}\bfseries,
  basicstyle=\footnotesize\color{black}\ttfamily,
  identifierstyle=\YAMLkeystyle,
  sensitive=false,
  commentstyle=\color{solarized@orange}\ttfamily,
  morecomment=[l]{\#},
  morecomment=[s]{/*}{*/},
  stringstyle=\YAMLstringstyle\ttfamily,
  moredelim=**[s][\YAMLkeystyle]{,}{:},   
  moredelim=**[l][\YAMLvaluestyle]{:},    
  morestring=[b]',
  morestring=[b]",
  literate =    {---}{{\ProcessThreeDashes}}3
                {>}{{\textcolor{solarized@red}\textgreater}}1
                {|}{{\textcolor{solarized@red}\textbar}}1
                {\ -\ }{{\mdseries\color{black}\ -\ \negmedspace}}3
                {\}}{{{\color{black} \}}}}1
                {\{}{{{\color{black} \{}}}1
                {[}{{{\color{black} [}}}1
                {]}{{{\color{black} ]}}}1
                {~}{\customtilde}1,
  breakindent=0pt,
  breakatwhitespace,
  columns=fullflexible
}

\lstdefinestyle{mathematica}
{
  language={Mathematica},
  basicstyle=\footnotesize\ttfamily,
  basewidth={0.53em,0.44em},
  numbers=none,
  tabsize=2,
  breaklines=true,
  escapeinside={@}{@},
  numberstyle=\tiny\color{black},
  showstringspaces=false,
  numberstyle=\tiny\color{solarized@base01},
  keywordstyle=\color{solarized@orange},
  stringstyle=\color{solarized@red}\ttfamily,
  identifierstyle=\color{solarized@orange}\ttfamily,
  commentstyle=\color{solarized@gray}\ttfamily,
  directivestyle=\color{solarized@orange}\ttfamily,
  emphstyle=\color{solarized@green},
  frame=single,
  rulecolor=\color{solarized@base2},
  rulesepcolor=\color{solarized@base2},
  literate={~} {\customtilde}1,
  moredelim=*[directive]\ \ \#,
  moredelim=*[directive]\ \ \ \ \#,
  mathescape=true
}
\lstset{style=python, breakatwhitespace}
\newcommand\textlst[1]{\lstinline!#1!}

\title{Geometry Optimization for Long-lived Particle Detectors}

\author{Thomas Gorordo}
\affiliation{Ernest Orlando Lawrence Berkeley National Laboratory, University of California, Berkeley, CA 94720, USA}

\author{Simon Knapen}
\affiliation{Ernest Orlando Lawrence Berkeley National Laboratory, University of California, Berkeley, CA 94720, USA}
\affiliation{Berkeley Center for Theoretical Physics, Department of Physics, University of California, Berkeley, CA 94720, USA}

\author{Benjamin Nachman}
\affiliation{Ernest Orlando Lawrence Berkeley National Laboratory, University of California, Berkeley, CA 94720, USA}

\author{Dean J.~Robinson}
\affiliation{Ernest Orlando Lawrence Berkeley National Laboratory, University of California, Berkeley, CA 94720, USA}
\affiliation{Berkeley Center for Theoretical Physics, Department of Physics, University of California, Berkeley, CA 94720, USA}

\author{Adi Suresh}
\affiliation{Ernest Orlando Lawrence Berkeley National Laboratory, University of California, Berkeley, CA 94720, USA}

\begin{abstract}
The proposed designs of many auxiliary long-lived particle (LLP) detectors at the LHC
call for the instrumentation of a large surface area inside the detector volume, 
in order to reliably reconstruct tracks and LLP decay vertices. 
Taking the CODEX-b detector as an example, 
we provide a proof-of-concept optimization analysis that demonstrates
the required instrumented surface area can be substantially reduced for many LLP models, while only marginally affecting the LLP signal efficiency.
This optimization permits a significant reduction in cost and installation time, and may also inform the installation order for modular detector elements.
We derive a branch-and-bound based optimization algorithm that permits highly computationally efficient determination of optimal detector configurations, 
subject to any specified LLP vertex and track reconstruction requirements.
We outline the features of a newly-developed generalized simulation framework, 
for the computation of LLP signal efficiencies across a range of LLP models and detector geometries.
\end{abstract}

\maketitle

\tableofcontents

\section{Introduction}
The LHC program is scheduled for ongoing upgrades and data collection until at least 2038. 
Given the relatively large luminosity that was already collected and correspondingly tightening limits, 
a potential discovery is increasingly likely to come from rare but striking classes of events, 
such as those involving new, long-lived particles (LLPs)~\cite{Beacham:2019nyx,Alimena:2019zri,Agrawal:2021dbo, Artuso:2022ouk,Bose:2022obr,Gori:2022vri}.
Compelling signatures along these lines are displaced decays-in-flight of exotic LLPs, 
which can generically arise in any theory containing a hierarchy of scales, 
technically natural small parameters, or loop-suppressed interactions.
LLPs are therefore ubiquitous in BSM scenarios~\cite{Curtin:2018mvb,Lee:2018pag, Aielli:2019ivi}.

Most relatively heavy---heavier than $\sim 10$\,GeV---LLPs may be searched for effectively at ATLAS, CMS and LHCb, even for relatively long lifetimes.
Searches for lighter LLPs are, however, often prohibitive, mainly because of large irreducible backgrounds and the corresponding trigger challenges.
An alternate LLP search paradigm therefore comprises looking for displaced decays-in-flight of exotic LLPs, 
in a displaced auxiliary detector that is well shielded 
(usually by a combination of passive and active shielding) from the interaction point (IP). 
While such `auxiliary' detectors would always have (much) smaller angular coverage than ATLAS and CMS, 
their increased shielding can more than compensate for their lower signal efficiency, 
by suppressing the backgrounds to negligible levels.
The detector transverse location moreover determines the type of physics to which one is sensitive: 
forward-located detectors are better suited for LLPs produced at low partonic center-of-mass energy~\cite{Feng:2022inv}, 
while more transverse detectors are required to find many types of LLPs produced in e.g.~Higgs decays or via other heavy mediators.

Typically, the design of transverse auxiliary LLP detectors requires not only a sizeable fiducial volume
but also a moderately large to large amount of instrumented surface area therein, for two reasons:
Firstly, in order to reliably reconstruct LLP decay vertices, each track should pass through multiple tracking stations to yield a large sample of hits.
The vertex resolution moreover typically degrades as the distance between the vertex and the first hit tracking station increases~\cite{Gligorov:2017nwh}.
Resistive plate chambers (RPCs) are a compelling choice for the tracking technology, because of their relatively high hit (and timing) resolution. 
Secondly, a more-or-less hermetic detector configuration is desirable to veto hard-cosmic, soft-cavern, and soft shielding-sourced backgrounds. 
However, since this type of background rejection coverage requires only a high detection efficiency and not the high hit resolution needed for track reconstruction of LLP decay products, 
substantially cheaper technologies can be considered, such as scintillators.

The baseline configuration for the \CODEXb detector,
proposed to be installed near LHCb's Interaction Point 8 (IP8)~\cite{Gligorov:2017nwh,Aielli:2019ivi}, 
comprises $\sim5100$\,m$^2$ of instrumented surface area in the form of triplets or sextets of RPC layers.
This configuration has already been shown in previous studies~\cite{Gligorov:2017nwh,Aielli:2019ivi} to have:
\textbf{(i)} a competitive sensitivity to a wide range BSM LLP scenarios, 
exceeding or complementary to the sensitivity of other existing or proposed detectors;
\textbf{(ii)} an achievable zero background environment, critical to LLP detection,
and accessible, well-equipped experimental location in the DELPHI/UXA cavern;
\textbf{(iii)} the ability to tag events within the LHCb detector, independently from the LHCb physics program;
\textbf{(iv)} a compact size, on detector scales, and consequently an expected modest cost.
A smaller proof-of-concept demonstrator detector, ``\CODEXbeta'', is proposed for operation during Run~3 of the LHC~\cite{Aielli:2019ivi}.

For this class of LLP detector, the required amount of RPC panels are the predominant driver of forecasted costs and installation time.  
It is therefore imperative to understand methods to optimize the design geometry of an LLP detector,
in order to minimize or substantially reduce the required amount of tracking versus background rejection surfaces, 
while maximizing (or preserving) the physics potential and sensitivity to LLP signals. 
Because of the broad range of BSM scenarios that may be probed by this class of auxiliary detector~\cite{Curtin:2018mvb, Aielli:2019ivi}, 
the optimization of the LLP detector geometry involves in turn the consideration of a broad range of well-motivated signal morphologies. 
These include many different signatures and kinematics, 
such that particle reconstruction requirements, efficiencies and acceptances vary widely.
For instance, the expected boost and rapidity distribution, as well as decay products of the LLP varies 
significantly between LLPs produced in Higgs decays versus hadron decays.
Further, in many well-motivated benchmark scenarios, the LLP may decay to various 
final states involving missing energy, photons, or a high multiplicity of softer tracks. 
These more complex decay morphologies can be much more challenging to detect and/or reconstruct.

In this work, we demonstrate the capabilities of a newly-developed versatile LLP simulation framework, 
that enables an understanding of the subtle interplay between the signal sensitivities of the detector geometry and the ability to probe underlying LLP models.
Using this framework, we simulate the response of variation in the detector geometry to different simulated BSM LLP production channels.
In Appendix~\ref{app:opt}, we derive in detail a `branch-and-bound'-type optimization algorithm, 
that permits highly computationally efficient determination of optimal detector configurations. 
These allow us to best allocate and arrange tracking stations in a particular LLP detector volume for any specified total instrumented surface area.
Importantly, the computation time for this algorithm scales approximately linearly with the number of tracking panels, 
rather than the exponential scaling one may naively expect from a brute-force method. 
(Details of the LLP simulation framework itself are provided in Appendix~\ref{app:simfr}.)
This algorithmic approach may be contrasted with other detector optimization techniques that use an automated optimizer,
either combinatorial or gradient-based as appropriate;
see e.g. Refs~\cite{Fanelli:2022rdm,modecn,Shirobokov:2020tjt}.

As a proof of concept, we apply this algorithm to a mildly simplified representation of the \CODEXb detector,
in order to demonstrate its capacity to identify optimized configurations.
We show that good sensitivity over the space of LLP scenarios can be attained for \CODEXb 
while reducing the amount of RPC layers by an $\mathcal{O}(1)$ fraction, correspondingly substantially reducing the forecast cost
as well as construction and installation times.
For example, for many LLP portals, we find optimized configurations 
in which the amount of tracking surface can be reduced by $50\%$ or more compared to baseline designs, 
while maintaining $80$\% or more relative efficiency for LLP detection compared to the \CODEXb baseline configuration.

This framework and optimization approach may be used in the future
to develop realistic optimized detector designs that incorporate buildability and other engineering constraints.
Furthermore, it may be exploited to further improve background rejection and reduce shielding costs: a subject of future work.
Apart from the baseline location proposed in Refs.~\cite{Gligorov:2017nwh,Aielli:2019ivi},
the framework may also be utilized to consider the sensitivities for other proposed \CODEXb locations near IP8,
and optimize their geometries. 
We do not consider other such locations in this work, however.
Similar applications are possible for other proposed auxiliary transverse LLP experiments, 
such as MATHUSLA~\cite{MATHUSLA:2022sze}, AL3X~\cite{Gligorov:2018vkc}, MAPP~\cite{Pinfold:2022quc} and ANUBIS~\cite{Bauer:2019vqk}.

\section{Detector and reconstruction requirements}
\label{sec:reqs}
In an LLP (or any) detector, the number of LLP events
$N_e = \varepsilon\, \mathcal{L}\, (\sigma \times \text{Br})$, where $\mathcal{L}$ is the integrated luminosity, 
$\sigma \times \text{Br}$ is the LLP production cross-section times branching ratio,
and $\varepsilon$ is the total signal efficiency (or just `total efficiency'), 
which is sensitive to the LLP boost and rapidity distribution, 
the LLP decay topology, the detector geometry, and reconstruction of the LLP decay vertex. 
The total efficiency itself may be factored into a volumetric efficiency $\varepsilon_v$, 
dependent only on the fiducial volume of the detector and the LLP proper decay length ($c\tau$),
and a reconstruction efficiency $\varepsilon_r$, 
that encodes the ability of the detector to reconstruct LLP decay tracks and the decay vertex.
Explicitly, on an event-by-event basis, one may define
\begin{equation} 
	\label{eq:weights}
	\varepsilon(\vec x, \vec p_i,\beta\gamma)\equiv\varepsilon_{v}(\vec x, \beta\gamma)\times\varepsilon_{r}(\vec x, \vec p_i)
\end{equation}
where $\vec x$ is the vertex position, $\beta\gamma$ the boost of the LLP and the $\vec p_i$ the momentum vectors of the decay products. 
The efficiency $\varepsilon_{v}(\vec x, \beta\gamma)$ represents the probability, 
under a particular normalization (see Appendix~\ref{app:norm}), that the decay occurred within the fiducial volume.
$\varepsilon_{r}(\vec x, \vec p_i)$ is the probability that the decay was reconstructed,
and depends directly on the detector configuration. 
The estimator of the total efficiency $\varepsilon$ is the sample average of the weights in Eq.~\eqref{eq:weights}, $\langle\varepsilon(\vec x, \vec p_i,\beta\gamma)\rangle$.
For a fixed volume, the reconstruction efficiency---and its relative ratio to a baseline configuration---is 
the key figure of merit in comparing the performance of various detector configurations. 
It is defined as the ratio of the total efficiency estimator over the estimator for the efficiency of a perfect detector, in other words 
\begin{equation}
	\label{eq:recodef}
	\varepsilon_{r} \equiv \langle\varepsilon(\vec x, \vec p_i,\beta\gamma)\rangle\big/\langle\varepsilon_v(\vec x, \beta\gamma)\rangle\,.
\end{equation}

Before turning to a study of these efficiencies,
we first present the requirements, specifications and performance for LLP reconstruction at \CODEXb, 
keeping in mind that these may be adapted to other LLP search environments.

\subsection{Baseline configuration and components}

The proposed \CODEXb detector technology follows the design for RPCs to be used in the ATLAS phase II muon chambers
(see Ref.~\cite{Collaboration:2285580} for technical descriptions).
The RPC tracking stations we consider will therefore be composed of $2\times 1$~m$^2$ triplet RPC layers.\footnote{
The more precise measurement for each chamber is $1.88 \times 1.03$\,m$^2$,
which we approximate by $2 \times 1$\,m$^2$ in this work for simplicity and clarity.}
Based on evolving technical specifications for the \CODEXbeta demonstrator detector (see also Ref.~\cite{Aielli:2019ivi}),
we assume the minimum detector element for the full \CODEXb detector will be a $2\times 2$ m$^2$ square ``panel'', 
containing two $2\times 1$\,m$^2$ RPC triplet layers.

\begin{figure}[bt]
	\begin{subfigure}[b]{0.3\textwidth}
        \centering
      	\includegraphics[width=5.5cm]{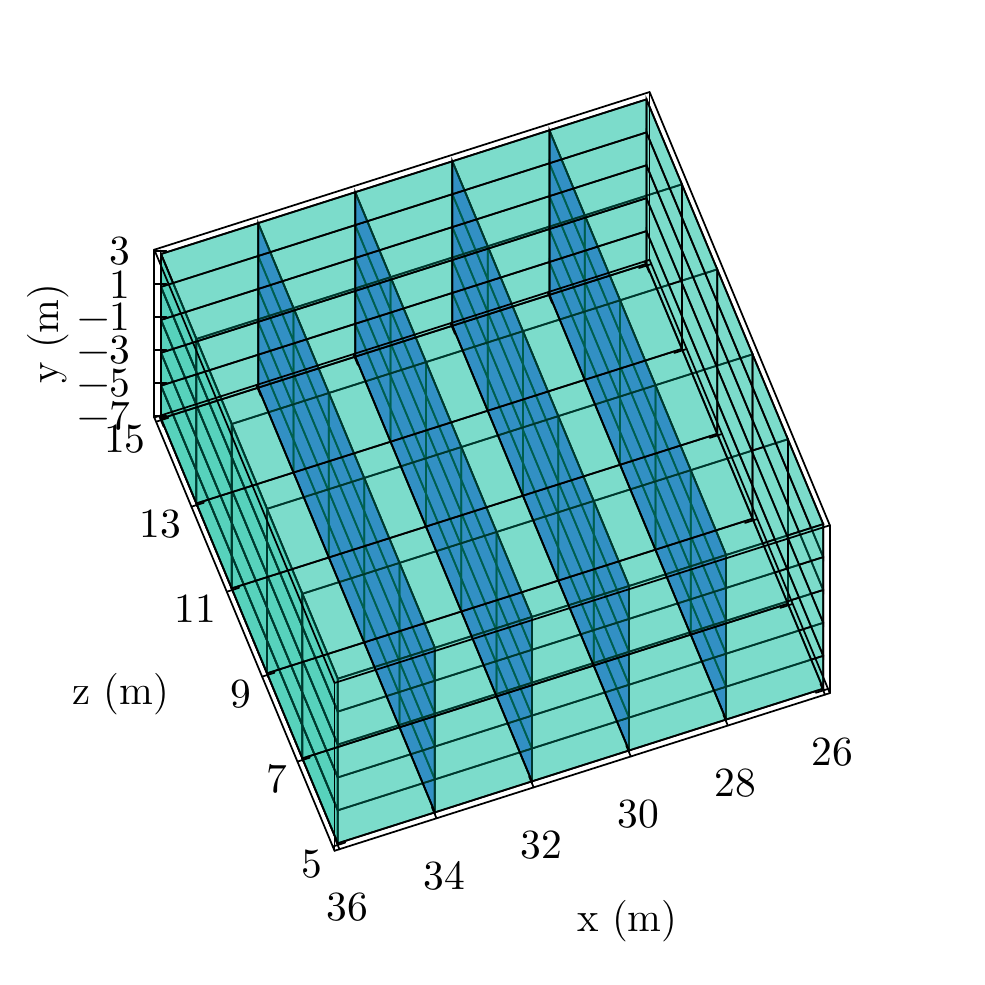}\hfil
        \caption{Baseline configuration\label{fig:base_geo}}
        \end{subfigure}\hfill
        \begin{subfigure}[b]{0.3\textwidth}
        \centering
      \includegraphics[width=5.5cm]{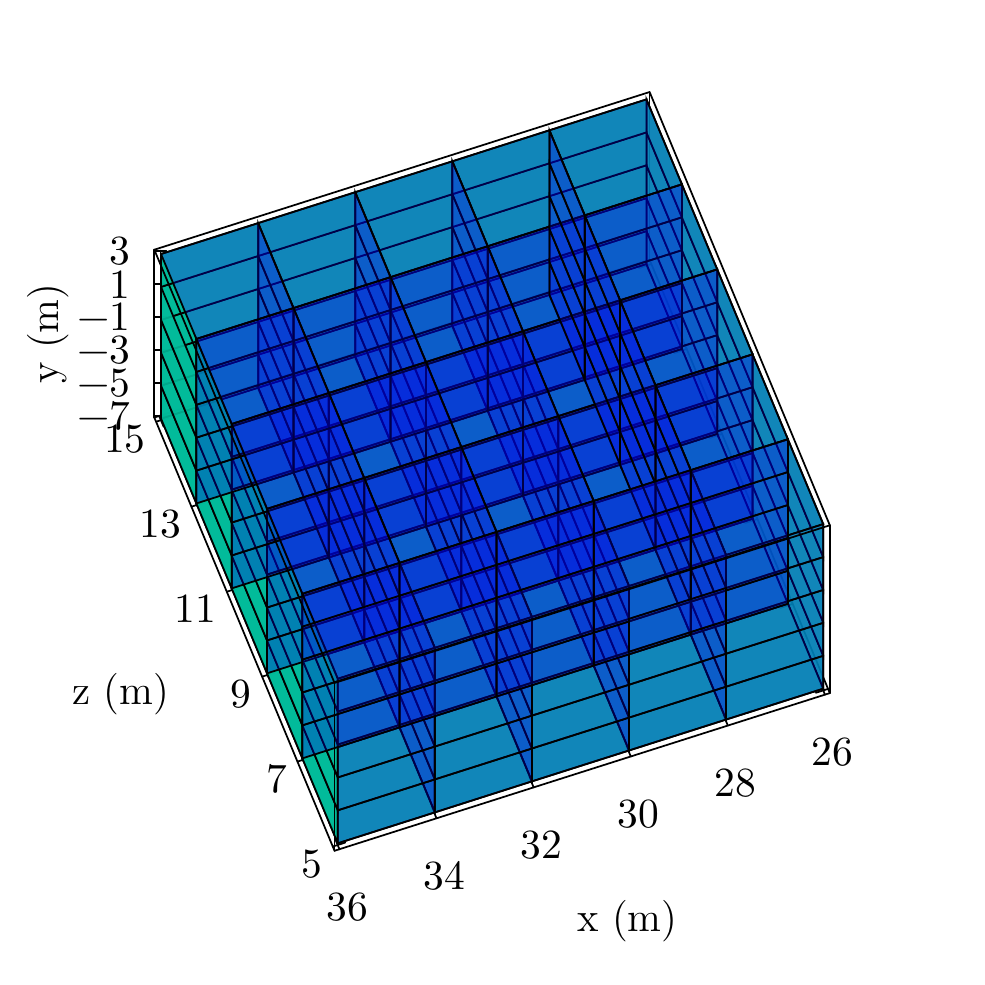}
        \caption{Envelope configuration\label{fig:full_geo}}
        \end{subfigure}\hfill
	\begin{subfigure}[b]{0.3\textwidth}
        \centering
      \includegraphics[width=5.5cm]{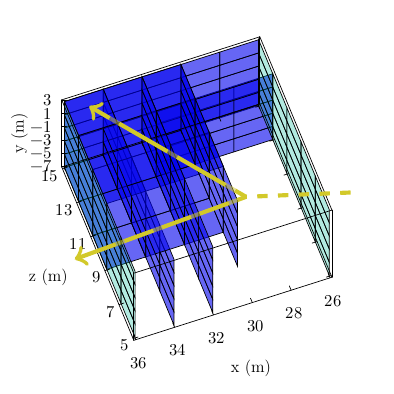}
        \caption{Heuristic configuration\label{fig:heur_geo}}
        \end{subfigure}
	\caption{\textbf{Left:} Schematic representation of the baseline detector geometry, based on Refs.~\cite{Gligorov:2017nwh,Aielli:2019ivi}, 
	with RPC sextets on the six external faces (green) and RPC triplets at four internal stations (blue).
	\textbf{Middle:} The maximal ``envelope'' detector geometry to be considered for optimization,
	with RPC sextets on four external faces (green) and RPC triplets at eight internal stations plus two external faces (blue).
	(Note this configuration does not fully contain the baseline.)
	\textbf{Right:} Heuristic partial configuration using $2\times2$~m$^2$ elements, with $\simeq50$\% of the instrumented surface 
	(excluding the $x=26$\,m sextet for background rejection), 
	but achieving $\sim50$--$80$\% relative efficiency depending on the LLP portal,
	compared to the baseline (see Sec.~\ref{sec:bench}). 
	In the rightmost figure we also show a schematic LLP decay topology, 
	in which a single LLP originating from the IP (dashed yellow) decays to two tracks (solid yellow), 
	each passing through multiple RPC triplet panels.}
\end{figure}

The baseline proposal for the \CODEXb detector comprises a $10\times 10 \times 10$\,m$^3$ fiducial volume 
located at $x=26$ to $36$\,m (transverse), $y=-7$ to $3$\,m (vertical) and $z = 5$ to $15$\,m (forward) with respect to IP8. 
The baseline detector design studied in Refs.~\cite{Gligorov:2017nwh,Aielli:2019ivi},
instrumented each of the six faces of this cubic detector volume with a sextet of RPC tracking layers, 
along with five triplets of RPC layers in the interior, uniformly spaced $1.67$\,m apart.
To adapt to the $2\times 1$\,m$^2$ element paradigm, we adjust the number of internal faces to four, 
as shown in Fig.~\ref{fig:base_geo}.
This corresponds to $400$ $2\times2$\,m$^2$ RPC triplet panels, 
such that the total RPC surface area required for this design is $\sim 4800$\,m$^2$.
The baseline design has the advantage of being relatively simple to simulate.
In Refs.~\cite{Gligorov:2017nwh, Aielli:2019ivi} it was shown that this design is capable of reconstructing
many LLP decay scenarios while achieving an $\mathcal{O}(1)$ tracking and reconstruction efficiency.

\subsection{Design drivers and detector configurations}
\label{sec:recoreq}

The design of the \CODEXb (or any LLP) detector is informed by four main considerations:

\begin{enumerate}[wide, labelwidth=!, labelindent=0pt, label = \quad \textbf{(\roman*)}, noitemsep, topsep =0pt]
\item \textbf{Acceptance}: 
Because the decay products of less boosted LLPs 
or high multiplicity multibody decays may be distributed over wide opening angles,
instrumentation is therefore required not just towards the back-end ($x = 36$\,m) face of the detector,
but also on the top, bottom and side faces of the cubic fiducial volume.

In the baseline design in Fig.~\ref{fig:base_geo}, all six external faces were instrumented equally, 
in order to ensure such softer LLP decays are captured.
However, even in this case,
typically one does not expect as high a density of hits on 
e.g.~the $z = 5$\,m face versus, say, the $z = 15$\,m face \cite{Aielli:2019ivi},
and it is unclear what fraction of the the $z = 15$\,m face 
needs to have a sextet layer for optimal sensitivity.
Because of this, we consider a maximal ``envelope'' detector configuration,
shown in Fig.~\ref{fig:full_geo}, 
in which the $x = 36$\,m, $z =15$\,m and $y = -7$\,m external faces have sextet panels, 
while $z = 5$\,m and $y = 3$\,m are triplet only. 
We stress that this configuration is merely meant as a maximal baseline, 
on which we shall apply our optimization procedure, 
and it is not intended \emph{per se} as a realistic configuration for the eventual \CODEXb detector.

\item \textbf{Vertex resolution}: 
Since no magnetic field will be available in \CODEXb, 
a good vertex resolution is essential to demonstrate the detection of an LLP.
A mentioned above, one important parameter in this respect is the distance from the LLP decay vertex to the first hit tracking layer~\cite{Gligorov:2017nwh}. 
For the baseline design, this motivated the inclusion of the five additional RPC triplet planes 
distributed uniformly in the interior of the fiducial volume along the $x$ direction, 
in order to achieve vertex resolutions at or below a centimeter.

For acceptable vertex reconstruction, in the study of the baseline design
it was required that: each track is composed of at least 6 hits;
and these hits are resolvable from hits generated by other tracks in the LLP decay.
In practice, assuming a $\mathcal{O}(\text{cm})$ spatial resolution for the RPC layers, 
the second condition implies a separation of $\gtrsim2$\,cm between resolvable hits on any given layer.
We retain this requirement in our optimization analyses.
Because, however, the present study is intended as a proof of concept for our optimization framework,
and because we wish to preserve intuitive insight into its results,
we make some simplifying assumptions concerning the vertex reconstruction: 
First, we assume the track six-hit condition is satisfied simply provided a track passes through two RPC triplet stations 
(in practice, the RPCs are obviously not perfectly efficient, requiring a correction).
Second, we do not impose constraints on the distance from the LLP decay vertex to the first hit layer, 
nor direct minimal requirements on the vertex resolution.
In other words, in this study we optimize for multitrack reconstruction efficiency, but not for vertex resolution.
Resolution requirements can, however, be straightforwardly imposed for a realistic optimization analysis of a full engineering design,
that also incorporates buildability and other engineering constraints.

In the baseline design, vertical internal layers lying in the $x$-$y$ plane were not considered, 
even though these may (or may not) more effectively provide coverage of the interior of the fiducial volume 
in light of these reconstruction requirements.
To examine their role under optimization, we include in the maximal envelope detector configuration 
an additional four internal faces uniformly spaced along the $z$ axis, containing RPC triplets,
as shown in Fig.~\ref{fig:full_geo}. 
The envelope configuration thus requires 450 panels,
corresponding to a total RPC surface area of $\sim 5400$\,m$^2$.

\item \textbf{Track momentum threshold}:
Because of the expected challenges in reconstructing multiply rescattered low momentum tracks, 
a minimum track threshold of $600$\,MeV was assumed for the baseline design study 
(for muon and pions to remain well inside the minimal ionizing particle regime through multiple detector elements; 
electrons instead undergo significant rescattering via Bremsstrahlung, though the same threshold was assumed).
For soft LLPs produced from (heavy) hadron decays, Ref.~\cite{Gligorov:2017nwh}, 
demonstrated that an $\mathcal{O}(1)$ reconstruction efficiency is maintained 
with a minimum track momentum threshold around $600$\,MeV or lower.

This threshold is the predominant driver of losses in reconstruction efficiency
for e.g.~the $b\to s S$ benchmark portal (see Sec.~\ref{sec:bkxportal}, below).
 At the same time, backgrounds produced by low energy (secondary) neutron scattering happen to grow significantly below this threshold~\cite{Aielli:2019ivi}. 
In this study, we retain the $600$\,MeV track momentum threshold as a track reconstruction constraint.

\item \textbf{Backgrounds}: Tracking stations on the front ($x = 26$\,m) face are needed to reject backgrounds 
from charged particles---primarily muons---emanating from the detector shielding.
For these stations, hit resolution is less 
important than detection efficiency, which may permit the use alternative technologies, such as scintillator planes: 
this requires further study by the \CODEXb collaboration, outside the scope of this work.
In all designs, we therefore always include a sextet of RPC layers on the front face, 
but we do not include this face---amounting to 50 $2\times2$\,m panels---when 
comparing the relative amount of instrumented surface for signal sensitivity.
That is, the instrumentation ratio for a configuration, with $n$ panels,
\begin{equation}
	r_n = (n- 50)/(N_\text{base} - 50)
\end{equation}
where $N_\text{base} = 400$, is the number of panels in the baseline configuration.

Soft backgrounds from cavern-based sources~\cite{Dey:2019vyo} and low-momentum neutral particles generated in the shielding
also motivate a hermetic coverage of the fiducial detector volume, in order to reject such backgrounds.
However, because once again detection efficiency is more crucial than hit resolution for the purpose of rejecting backgrounds, 
one may contemplate replacing the RPC triplet panels with scintillator on the outer faces of the detector, 
wherever the former would provide subleading sensitivity to LLP signal detection and reconstruction.
Characterizing the degree to which this would degrade the signal sensitivity is a main goal of the present work.

In order to provide an intuitive comparison with the baseline design shown in Fig.~\ref{fig:base_geo},
as we discuss the performance of the model benchmarks chosen below
we also consider the performance of a `heuristic' configuration as shown in Fig.~\ref{fig:heur_geo}. 
This configuration is an informed guess for an optimized configuration, 
that attempts to place instrumented surface only in locations expected to experience higher numbers of hits, 
while likely still satisfying the reconstruction requirements:
It has a sextet only on the back $x =36$\,m face (and the $x =26$\,m face) 
with $r_n \simeq 0.51$, and many entire or partial faces have been removed.
As we shall see below in Sec.~\ref{sec:bench}, this configuration nonetheless has a relative reconstruction 
efficiency ranging from $50$--$80$\% of the baseline performance,
depending on the LLP production and decay portals.
We will find, however, that optimized configurations subject to the above reconstruction requirements will outperform the heuristic configuration.

\end{enumerate}

\section{Benchmark topologies and simplified models}
\label{sec:bench}

Exotic LLPs may generically arise in any theory containing a hierarchy of scales, technically natural small parameters, or loop-order suppressions,
and are therefore ubiquitous in BSM scenarios~\cite{Curtin:2018mvb,Aielli:2019ivi}. 
An analogous SM phenomenology is found in the long-lived $K^0_L$, $\pi^\pm$, neutron and muon. 
In this optimization study, we will limit ourselves to three combinations of simple LLP production and decay portals, 
that act as good representatives for the typical event topologies that may be searched for at LLP detectors such as \CODEXb.
These are summarized in Table~\ref{tab:topos}, and discussed in more detail in the following subsections.
Also indicated in Table~\ref{tab:topos} are other commonly considered LLP models,
in particular axion-like particles (ALPs) and heavy neutral leptons (HNLs), whose typical production and decay morphologies are analogous to the considered models.
For each production and decay portal, we survey several LLP masses in the $0.5$--$10$\,GeV range, 
leading to a total of eleven benchmark models.

\begin{figure}[t]
	\includegraphics[width=0.4\linewidth]{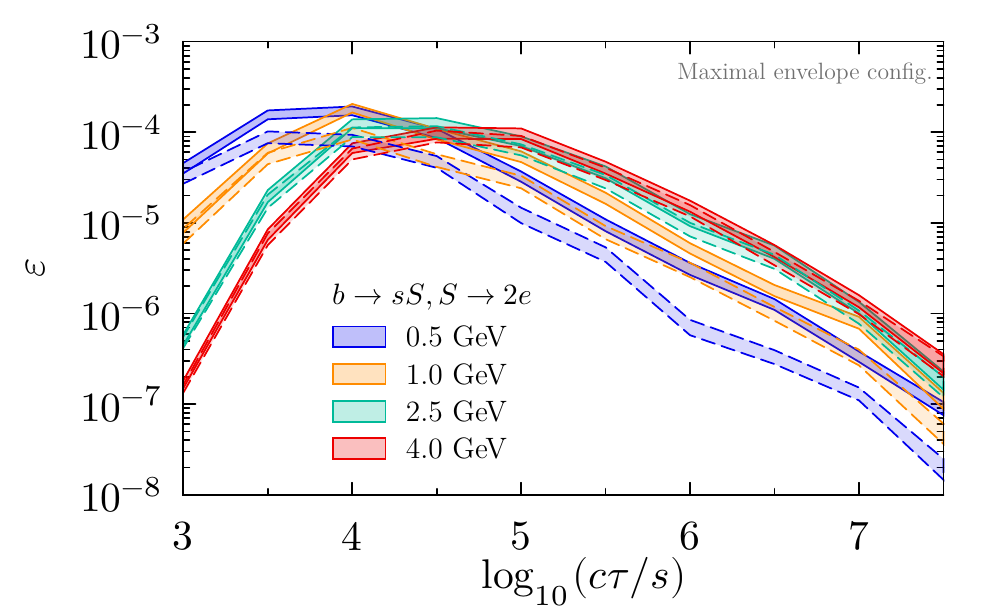}\hfil
	\includegraphics[width=0.4\linewidth]{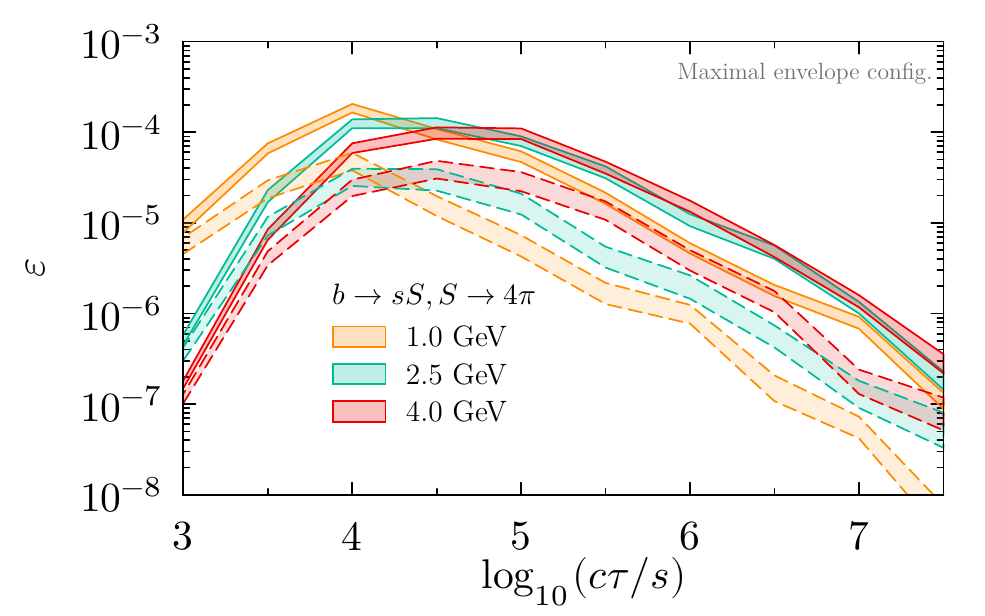}\\
	\includegraphics[width=0.4\linewidth]{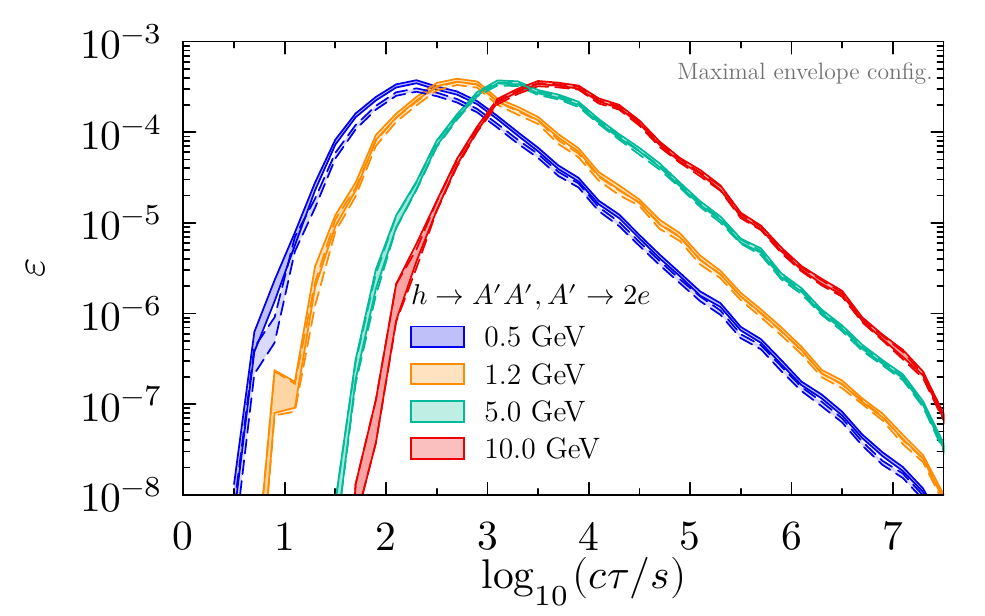}	
	\caption{The $1\sigma$ CL total efficiency bands for the full nominal configuration from raw MC simulation---i.e.
	without post-simulation power-law fitting---as 
	a function of the LLP lifetime $c\tau$, excluding (solid) and including (dashed)
	track reconstruction requirements.
	All uncertainties are from MC statistics.}
	\label{fig:abseffs}
\end{figure}

We simulate total efficiencies and vertex reconstruction efficiencies using MC samples generated with \texttt{Pythia 8} \cite{Sjostrand:2014zea}, 
that are then reweighted and further processed with a dedicated, generalized LLP simulation framework:
The details of this simulation framework are discussed in Appendix~\ref{app:simfr}.
Our benchmark models will also serve as good examples of use cases that determine its required specifications and capabilities.
In Fig.~\ref{fig:abseffs} we show the total efficiencies for the various benchmarks, as a function of $c\tau$.
The displayed efficiencies are from raw simulation, 
without fitting to the known linear powerlaw behavior in the long-lifetime regime,
and we include the uncertainties from MC statistics. 

An important takeaway is that the impact of the reconstruction requirements is roughly a constant factor across broad ranges of $c\tau$.
(This is expected in the long lifetime regime,
in which the LLP characteristic displacement from the IP, $\beta \gamma c \tau$, is much greater than the detector displacement.
This is because the volumetric efficiency, which is determined by probability of decay, scales as 
$e^{-|\vec x|/\beta\gamma c\tau}/\beta\gamma c\tau
\simeq 1/\beta\gamma c\tau$ for $|\vec x| \ll \beta\gamma c\tau$. In this regime the dependence on $c\tau$ therefore drops out in Eq.~\eqref{eq:recodef}.) 
As a result, even though the LLP lifetime is formally an independent parameter of the benchmark models, 
we shall hereafter discuss and compute vertex reconstruction efficiencies by marginalizing over a $c\tau$ range close to the peak,
at which the kinematic distributions should well represent those for the entire long-lifetime regime.
We do not optimize for LLPs in the short-lifetime regime, where existing LHC experiments are expected to set the most competitive limits.

\begin{table}
\renewcommand*{\arraystretch}{0.9}
\newcolumntype{C}{ >{\centering\arraybackslash } m{3cm} <{}}
\newcolumntype{D}{ >{\centering\arraybackslash } m{1.5cm} <{}}
\scalebox{0.7}{\parbox{1.4\linewidth}{
\begin{tabular*}{\linewidth}{@{\extracolsep{\fill}}CDDCCCCC}
	\hline
	Portal/Model & \multicolumn{2}{c}{LLP production}  & \makecell{Finite \\ hit resolution} & \makecell{Track momentum \\ threshold} & \makecell{Acoplanar \\topology} & Masses 
	& \multirow{2}{*}[3pt]{\makecell{Analogous \\ simplified \\ models}} \\
	&single & multiple & (small opening angle) & (soft tracks) & (missing energy) &  [GeV] & \\
	\hline\hline
	\makecell{\\$h \to A' A'$ \\[0pt]  [$A' \to 2e$]} & --- & \checkmark & \checkmark & --- & --- & $0.5$, $1.2$, $5$, $10$ & ALP \\	
	\makecell{\\$b \to s S$ \\[0pt]  [$S \to 2e$]} & \checkmark & --- & --- & \checkmark & ---  & $0.5$, $1$, $2.5$, $4$ & ALP\\
	\makecell{\\$b \to s S$ \\[0pt]  [$S \to 4\pi$]} & \checkmark & --- & --- & \checkmark & \checkmark & $1$, $2.5$, $4$ & ALP, HNL\\	
	\hline	
\end{tabular*}
}}
\renewcommand*{\arraystretch}{1.}
\caption{The three benchmark production and decay portals, their topologies and typical sensitivity-limiting features,
along with the selected mass points for each portal, for a total of eleven benchmark models.
Also indicated are other simplified LLP models---axion-like particles (ALP) and heavy neutral leptons (HNL)---that can have 
analogous morphologies to particular combinations of production and decay portals.}
\label{tab:topos}
\end{table}

\subsection{$h \to A' A'$, $A' \to 2e$}

The first topology comprises LLP pair production at high boost, with each LLP decaying to two tracks.
In particular, we consider LLP production via an exotic decay of the SM Higgs, 
in which it decays to a pair of highly-boosted GeV scale LLPs.
There are many ways of realizing this scenario in a specific model: 
generally the branching ratio of the Higgs decay to the LLPs is independent from the LLP lifetime.
For concreteness, we take the example of a dark photon ($A'$) LLP~\cite{Schabinger:2005ei,Gopalakrishna:2008dv,Curtin:2014cca}, 
treating $\mathrm{Br}[h \to A' A']$, the LLP lifetime, $c\tau$, and its mass, $m_{A'}$, as independent parameters. 
In the $\sim$ GeV mass range we consider, 
the $A'$ itself predominantly decays to pairs of charged leptons or hadrons~\cite{Meade:2009rb,Buschmann:2015awa}:
for simplicity we consider $A'$ decays just to electron positron pairs.
 
In Table~\ref{tab:trkreceff} we show the vertex reconstruction efficiencies
for various benchmarks involving different LLP masses and production or decay portals.
The vertex reconstruction efficiencies for this portal are in the $\varepsilon_r(\text{base})$, $\varepsilon_r(\text{env}) \sim 80$--$90$\% range 
for both the baseline and maximal envelope configurations,
with the main reductions arising from $\gtrsim$ 2\,cm hit separation criterion.
For the heuristic configuration, the efficiency lies in the $\varepsilon_r(\text{heur}) \sim70$-$80$\% range
even though this configuration features approximately half the amount of RPC surface area.
This observation raises the prospect, which we confirm below, that a more systematic optimization procedure might permit 
further reductions in the instrumentation ratio, $r_n < r_{n(\text{heur})}$, 
while maintaining competitive sensitivities.

\begin{table}[t]
\renewcommand*{\arraystretch}{1.5}
\newcolumntype{C}{ >{\centering\arraybackslash $} m{2cm} <{$}}
\newcolumntype{D}{ >{\centering\arraybackslash $} l <{$}}
\newcolumntype{E}{ >{\centering\arraybackslash $} c <{$}}
\scalebox{0.7}{\parbox{1.4\linewidth}{
\begin{tabular*}{\linewidth}{ECCC|ECCC|ECCC}
	\hline
	\multicolumn{4}{c|}{$h \to A' A'~(2e)$ } &  \multicolumn{4}{c|}{$b \to sS~(2e)$} & \multicolumn{4}{c}{$b \to sS~(4\pi)$} \\[10pt]
	\multirow{-2}{*}{\makecell{$m_{A'}$ \\[5pt] [GeV]}}  & \text{Baseline} & \text{Envelope} & \text{Heuristic} & 
	\multirow{-2}{*}{\makecell{$m_{S}$ \\[5pt] [GeV]}}  & \text{Baseline} & \text{Envelope} & \text{Heuristic} & 
	\multirow{-2}{*}{\makecell{$m_{S}$ \\[5pt] [GeV]}}  & \text{Baseline} & \text{Envelope} & \text{Heuristic}\\ 
	  \hline\hline
	  0.5 & 0.83(3) & 0.82(3) & 0.72(3) & 0.5 & 0.53(5) & 0.52(4) & 0.41(4) & \text{---} & \text{---} & \text{---} &  \text{---}\\ 
	1.2 & 0.93(4) & 0.91(4) & 0.79(3) & 1.0 & 0.58(5) & 0.56(5) & 0.35(3) & 1.0 & 0.24(3) & 0.23(3) & 0.18(2) \\ 
	5.0 & 0.97(4) & 0.95(4) & 0.80(3) & 2.5 & 0.92(8) & 0.81(7) & 0.45(4) & 2.5 & 0.26(3) & 0.24(3) & 0.16(2) \\ 
	10.0 & 0.97(4) & 0.96(4) & 0.77(3) & 4.0 & 0.99(9) & 0.87(8) & 0.47(5) & 4.0 & 0.37(4) & 0.35(4) & 0.22(3) \\ 
	\hline
\end{tabular*}
}}
\caption{
Reconstruction efficiencies, $\varepsilon_r$, marginalized over $c\tau$, 
requiring six hits per track and a minimum track momentum threshold of $600$\,MeV.
All uncertainties (shown in parentheses) are from MC statistics.}
\label{tab:trkreceff}
\renewcommand*{\arraystretch}{1}
\end{table}

\subsection{$b \to s S$, $S \to 2e$}
\label{sec:bkxportal}

The second topology comprises LLP production at moderate boost, with the LLP decaying to two tracks.
In particular, we consider an exotic $b$-hadron decay, $b \to s S$. 
The particle $S$ is a light, new singlet added to the SM through the Higgs-mixing portal
\begin{equation}\label{eq:lag}
	\mathcal{L}\supset \mu\, S H^\dagger H\quad\text{with} \quad \mu \ll m_W.
\end{equation}
In this scenario, one benefits from the huge $b$-hadron production cross section, 
and very narrow decay width of the $B_{u,d,s}$ or $\Lambda_b$ ground state hadrons. 
In the mass basis, $S$ has a Yukawa coupling to all SM fermions proportional to their masses and the mixing angle, $\sin\theta$, between $S$ and the SM Higgs. 
The $B\to X_s S$ inclusive decay, as a representative of all inclusive $b$-hadron decays, 
occurs through an electroweak penguin with branching ratio 
$\text{Br}[B\to X_s \phi]\approx 6.2 \times \sin^2\theta$~\cite{PhysRevD.26.3287,Chivukula:1988lo,Grinstein:1988yu}. 
The lifetime is inversely proportional to $\sin^2\theta$ 
and depends further on a somewhat complicated function of $m_S$: 
See \cite{Winkler:2018qyg,Gershtein:2020mwi} for the most recent calculation.
In the $\sim$ GeV mass range we consider, the branching ratios for $S$ decays to pairs of charged leptons  or hadrons are $\mathcal{O}(1)$ (See e.g.~\cite{Gunion:1989we}). 
Once again for simplicity we consider $S$ decays just to electron positron pairs.

From Table~\ref{tab:trkreceff}, the vertex reconstruction efficiencies for this portal 
lie in the $\varepsilon_r(\text{base})$, $\varepsilon_r(\text{env}) \sim 50$--$90$\% range for the baseline and full nominal configurations,
with the main reductions arising from the $600$\,MeV track momentum threshold.
For the heuristic configuration, the efficiency lies in the $\sim50$\% range,
such that for some lighter mass points, the heuristic configuration performs 
comparably to the other two,
while for heavier LLPs the sensitivity roughly scales with the instrumentation ratio $r_{n(\text{heur})}$.

\subsection{$b \to s S$, $S \to 4\pi$}

Beyond the case that the LLP decays to two (or more) tracks, a wide range of LLP scenarios may involve decays of the LLP to tracks plus missing energy. 
For instance, HNLs may feature decays to final states containing neutrinos.
Other LLPs may feature decays to long-lived neutral hadrons, such as the $K_L^0$, that may escape the detector,
or to neutral hadrons with rapid electromagnetic decays, such as the $\pi^0$, 
that cannot be reconstructed without a calorimeter (which we assume the LLP detector does not have).
To capture this topology, we also consider the above Higgs mixing portal, but with $S$ decaying to $\pi^+\pi^-\pi^0\pi^0$,
in which the $\pi^0$ is treated as missing energy.

Compared to the $b \to s S$, $S \to 2e$ mass benchmarks,
the typically softer tracks in the $S \to 4\pi$ decay mode lead to further
reductions in the vertex reconstruction efficiency,
because of the $600$\,MeV track momentum threshold.
We see in Table~\ref{tab:trkreceff}, however, that just as for the $S \to 2e$ mode, 
for the lighter $1$\,GeV mass point, 
the heuristic configuration performs comparably to the other two,
while for heavier LLPs the sensitivity begins to scale as $r_{n(\text{heur})}$. 

\section{Estimators and optimization}

\subsection{General approach}

In order to optimize the detector geometry, 
one seeks to understand which subset(s) of RPC panels exhibit 
greater or lesser sensitivity to our selection of LLP benchmark models.
In particular, for the maximal envelope configuration $\Sigma = \{p_i\}_{i \le N}$ of $N$ total panels, 
the goal is to understand which subset of size $n \le N$ panels, $\sigma^{(n)} \subset \Sigma$,
maximizes the relative vertex reconstruction efficiency
\begin{equation}
	\rho_{n,b}= \varepsilon_{r,b}(\sigma^{(n)})/\varepsilon_{r,b}(\text{base})\,,
\end{equation}
in which $\varepsilon_{r,b}(\sigma)$ 
denotes the tracking reconstruction efficiency of configuration $\sigma$ for benchmark $b$, 
as shown in Table~\ref{tab:trkreceff}.
We use here and hereafter a superscript in parentheses to denote the size---the cardinality---of a subset, so that $|\sigma^{(n)}| = n$.
In general, the optimal configuration of size $n$, denoted $\sigma^{(n)}_{\text{opt}}$, 
will be sensitive to the particular benchmark under consideration. 
We will therefore seek maximizations that perform best on average over the space of selected benchmarks, 
assuming a uniform prior for the weight of any LLP benchmark versus the others. 
That is, we will seek $\sigmaopt{n}$ that optimizes the objective function $f = \sum_b\rho_{n,b}$.\footnote{
In principle, the optimization problem may also admit relative panel weights, 
that encode some relative ``cost'' between different panels, and require their sum to be bounded.
The following discussion may be straightforwardly generalized to include such considerations.}

If the objective function, $f$, were additive---i.e. $f(\sigma \cup \sigma') = f(\sigma) + f(\sigma')$---then 
the optimization problem is a special case of the classic ``0/1-knapsack problem",
whose solution can be found straightforwardly from a ``greedy'' algorithm.
This proceeds as:
(i) determine the values of $f(\{p_i\})$ for each individual panel;
(ii) order the panels accordingly from highest to lowest value; and 
(iii) take the $n$ highest-valued panels.
Put a different way, sorting by individual objective function values yields an
ordering of panels $\{p_{z_i}\}$ such that $f(\{p_{z_1}\}) \ge f(\{p_{z_2}\}) \ge....f(\{p_{z_N}\})$,
and the solutions are $\sigmaopt{n} = \{p_{z_i}\}_{i \le n}$ for each $n \le N$.

\begin{figure}[t]
\includegraphics{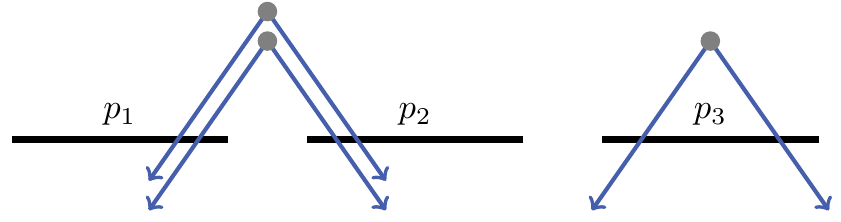}
\caption{A configuration of three RPC triplet panels, $\Sigma = \{p_1,p_2,p_3\}$, subject to three LLP decay events, each with two tracks (blue). 
Assuming in this example only that just three hits per track---i.e. passing through one triplet---are required for reconstruction,
but still retaining the requirement that both tracks are reconstructed,
the single panel subset with highest reconstruction efficiency is $\sigmaopt{1} = \{p_3\}$.
The two-panel subset, however, with highest efficiency $\sigmaopt{2} = \{p_1,p_2\} \not\supset \sigmaopt{1}$.}
\label{fig:exord}
\end{figure}

\begin{figure}[t]
\includegraphics{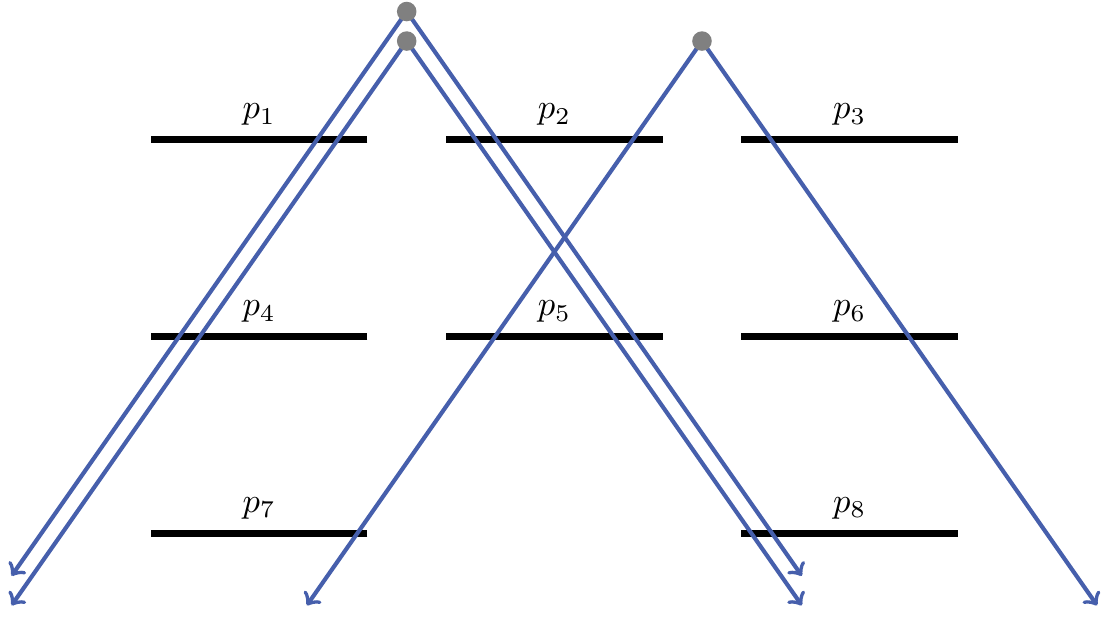}
\caption{A configuration of eight RPC triplet panels subject to three LLP decay events, each with two tracks (blue). 
The reconstruction requirement that both tracks in an event have a minimum of $2$ triplet hits each (six hits per track)
leads to degeneracies for the vertex reconstruction efficiencies of various configurations:
The union of $\{p_1, p_4\}$ with any two of $p_{2,5,8}$ all have $\varepsilon_r = 2/3$, 
even though $p_{2,5}$ have higher numbers of hits than $p_8$;
if $p_7$ must be included in a configuration, 
then the union $\{p_1, p_3, p_4,p_6,p_7\}$ with any two of $\{p_2, p_5, p_8\}$
has $\varepsilon_r = 1$.
 }
\label{fig:exord2}
\end{figure}

For the tracking and vertex reconstruction requirements we consider in Sec.~\ref{sec:recoreq}, 
however, the objective function, $f = \sum_b\rho_{n,b}$, is
manifestly nonadditive, because of the effects of hit-hit correlations in each event.
This can be seen for the example shown in Fig.~\ref{fig:exord}:
$\varepsilon_r({\{p_{1}\}}) = \varepsilon_r({\{p_{2}\}}) = 0$, but $\varepsilon_r({\{p_{1}, p_{2}\}}) = 2/3$ 
(assuming in this example only three hits per track---i.e. passing through one triplet---is required for reconstruction).
An additional complication arises when we consider that decay events may 
often saturate the minimum six hits per track requirement (corresponding to hitting at least two RPC triplet layers).
As shown in Fig.~\ref{fig:exord2}, 
a minimum number of hits per track can lead to degeneracies in the vertex reconstruction efficiency among various configurations,
such that the presence of some panels are anticorrelated with others, for a fixed cardinality $n$.
Put in other words, such requirements can lead to panel-panel redundancies.
As a result, the class of problem we must consider involves an objective function for which the difference
\begin{equation}\label{eq:diff_objective}
	 f(\sigma \cup \sigma') - f(\sigma) - f(\sigma')\,,
\end{equation}
may take positive or negative values,
because of the effect of hit-hit correlations and panel-panel redundancies, respectively:
The set function $f$ is neither superadditive nor subadditive.

Because of the hit-hit correlations and panel-panel redundancies in each event, 
in general for $n' < n$, the globally optimal subset 
$\sigmaopt{n'}$ need not be a subset of the globally optimal $\sigmaopt{n}$.
(See Fig.~\ref{fig:exord} for a trivial example.) 
We are interested here, however, in a prioritization of the RPC panels by which one may incrementally install an experiment, 
and thus further panels may be added but not removed from a given configuration. 
We therefore seek an \emph{optimized ordering} of configurations,
in which each $\sigmaopt{n}$ contains 
$\sigmaopt{n'}$ for $n' \le n$, i.e.
\begin{equation}
	\label{eq:chainofopt}
	\varnothing = \sigmaopt{0}\subset\sigmaopt{n_1}\subset\ldots
	\subset\sigmaopt{n_k}\subset\ldots
	\sigmaopt{n_{K-1}}\subset\sigmaopt{n_K=N} = \Sigma\,.
\end{equation}
This nesting requirement means that it may not be possible to construct a globally optimal $\sigmaopt{n}$ for every $n \le N$, 
i.e. such that Eq.~\eqref{eq:chainofopt} is satisfied,
because it may happen that one would need to remove panels from the existing $n$-panel configuration and add others elsewhere to achieve a global optimum for $n+1$
(see e.g. Fig.~\ref{fig:exord}).
Instead, we will show that it is possible to construct an optimized ordering for a (large) subset of cardinalities $n_k \le N$, 
such that $n_{k+1} - n_k \sim \text{few}$, and interpolate for $n_k \le n \le n_{k+1}$:
we refer to this as an \emph{optimized partial ordering}.

\subsection{Hit-weight estimator}
Before proceeding to the implementation of a systematic optimization, it is 
useful to first gain intuition for the general behavior and anticipated 
characteristics of $\rho_n[b]$ as a function of $n$. To this end, one can 
examine the ordering determined by the number of (weighted) hits on each of the
RPC panels belonging to the maximal envelope configuration.
Specifically, for benchmark $b$, the cumulative hit weight for the 
$i$th panel 
\begin{equation}
	h_{i,b} = \sum_{\alpha \in E_{i,b}} w_{\alpha,b}\,,
\end{equation}
where $E_{i,b}$ is the set of events in benchmark $b$ whose tracks hit the 
$i$th panel (possibly multiple times, if multiple tracks hit the panel), subject
to the pertinent track reconstruction requirements, and $w_{\alpha,b}$ is the
weight of the event to which the track belongs.
The latter is defined as the event MC generation weight multiplied by the LLP decay probability (see also App.~\ref{app:norm}).  
Normalizing over the space of panels and imposing a 
uniform prior over the set of benchmarks, leads to the benchmark-averaged hit 
weight
\begin{equation}
	\label{eqn:hwe}
	\hat{h}_i= \frac{1}{n_B}\sum_b \Big[ h_{i,b}\Big/\sum_j h_{j,b} \Big]\,,
\end{equation}
where $n_B$ is the number of benchmarks. This weight may be thought of as the 
relative probability that a panel is hit by an LLP decay track. 
Sorting the panels from highest to lowest $\hat h_i$ produces an ordering.
Although it neglects the effects of hit-hit correlations from either the same 
track or another track in the same event, one might expect this ordering to act
as a naive estimator for an optimized ordering, whose efficiency provides a lower bound for the latter.
Further, one might expect an estimator based on hit weight to be correlated with decreasing distance from the LLP vertex 
(and thus correlated with good LLP vertex resolution)
as the farther a panel is located from a vertex, the fewer tracks are expected to pass through it per event.

\begin{figure}[t]
	\includegraphics[width=0.4\linewidth]{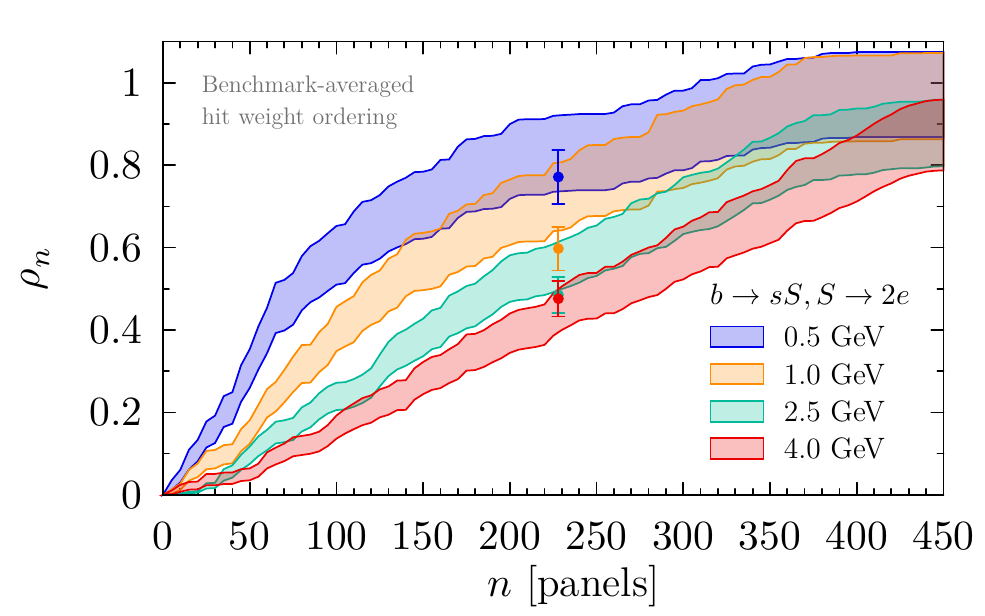}\hfil
	\includegraphics[width=0.4\linewidth]{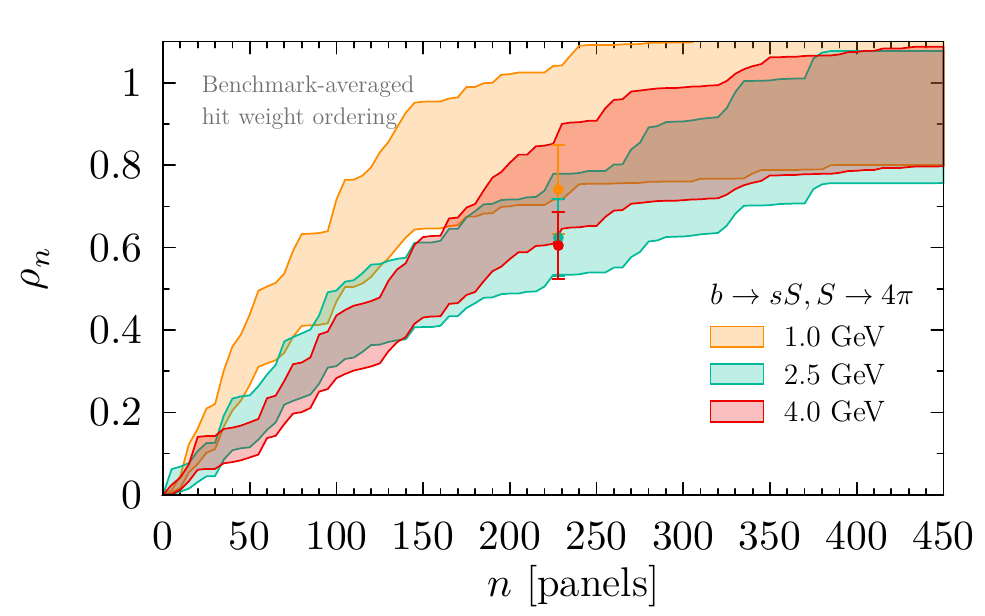}\\
	\includegraphics[width=0.4\linewidth]{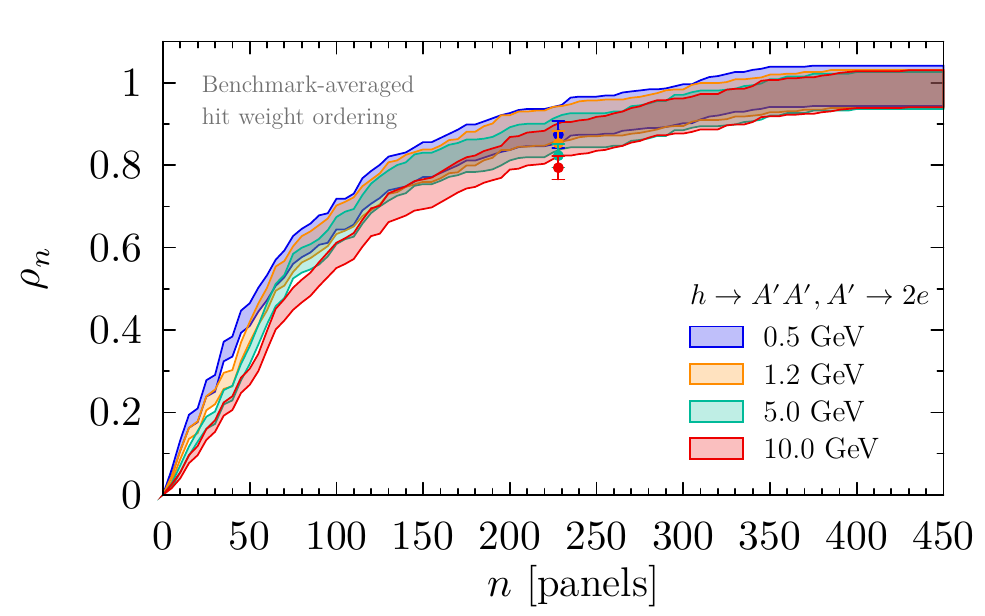}	
	\caption{The $1\sigma$ CL bands for relative vertex reconstruction efficiencies as a function of number of panels
	selected from the maximal envelope configuration,
	as determined by an ordering 
	using the benchmark-averaged hit weight.
	All uncertainties are from MC statistics.
	Also shown by correspondingly colored single data points are the relative vertex reconstruction efficiencies of the heuristic configuration of Fig.~\ref{fig:heur_geo}.}
	\label{fig:naiveordering}
\end{figure}

In Fig.~\ref{fig:naiveordering} we show the relative reconstruction 
efficiencies $\rho_n$ for the various different benchmarks listed in 
Table~\ref{tab:trkreceff}, as a function of the number of RPC panels selected from the maximal envelope configuration, 
with ordering determined by $\hat{h}_i$.
Of particular importance is the relatively steep negative curvature for some of
the benchmark models: the relative vertex reconstruction efficiency for these 
rises quite quickly as a function of the number of panels. This demonstrates 
that for some benchmarks, there exist configurations with high values for 
$\rho_n$ for relatively low $n$, even using this naive hit weight 
estimator.
These orderings achieve a comparable or slightly higher relative efficiency versus that of the heuristic configuration of Fig.~\ref{fig:heur_geo},
whose relative efficiency is shown by colored single data points in Fig.~\ref{fig:naiveordering}. 

\begin{figure}[bt]
      	\includegraphics[width=7cm]{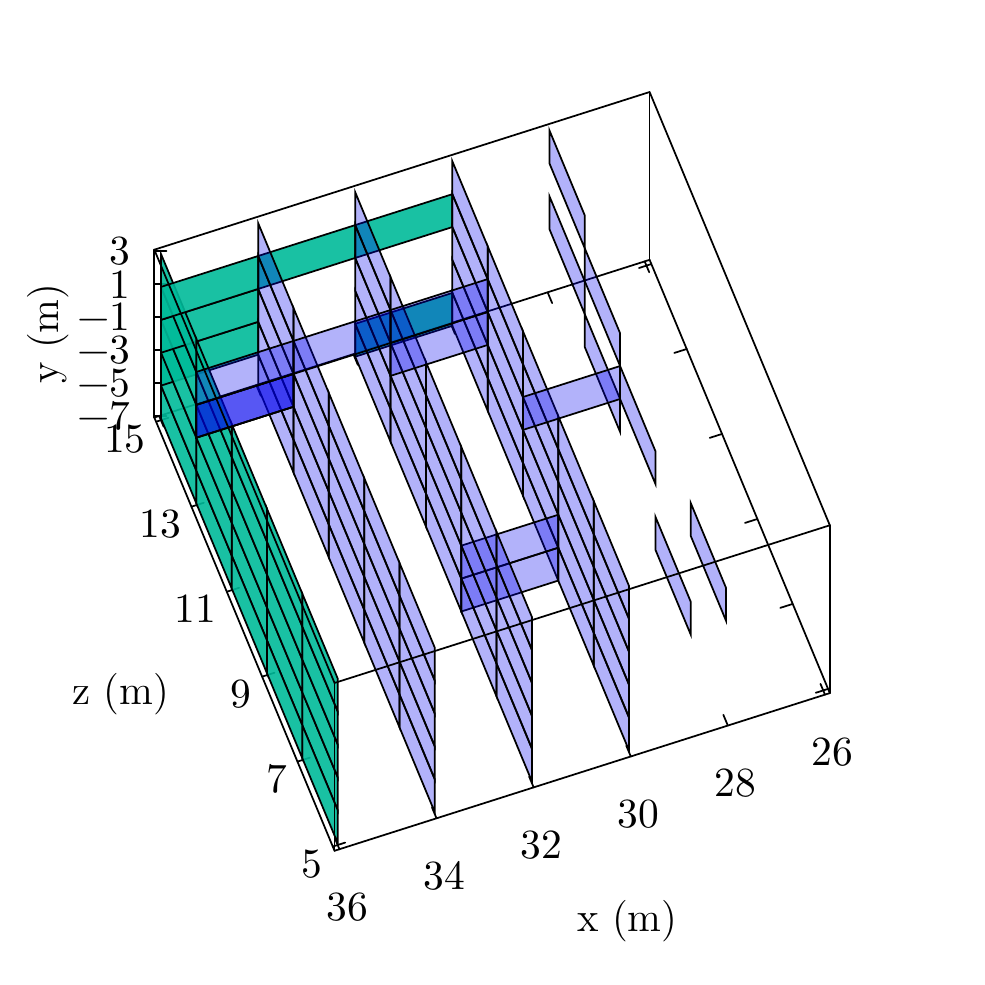}\hfil
	\includegraphics[width=7cm]{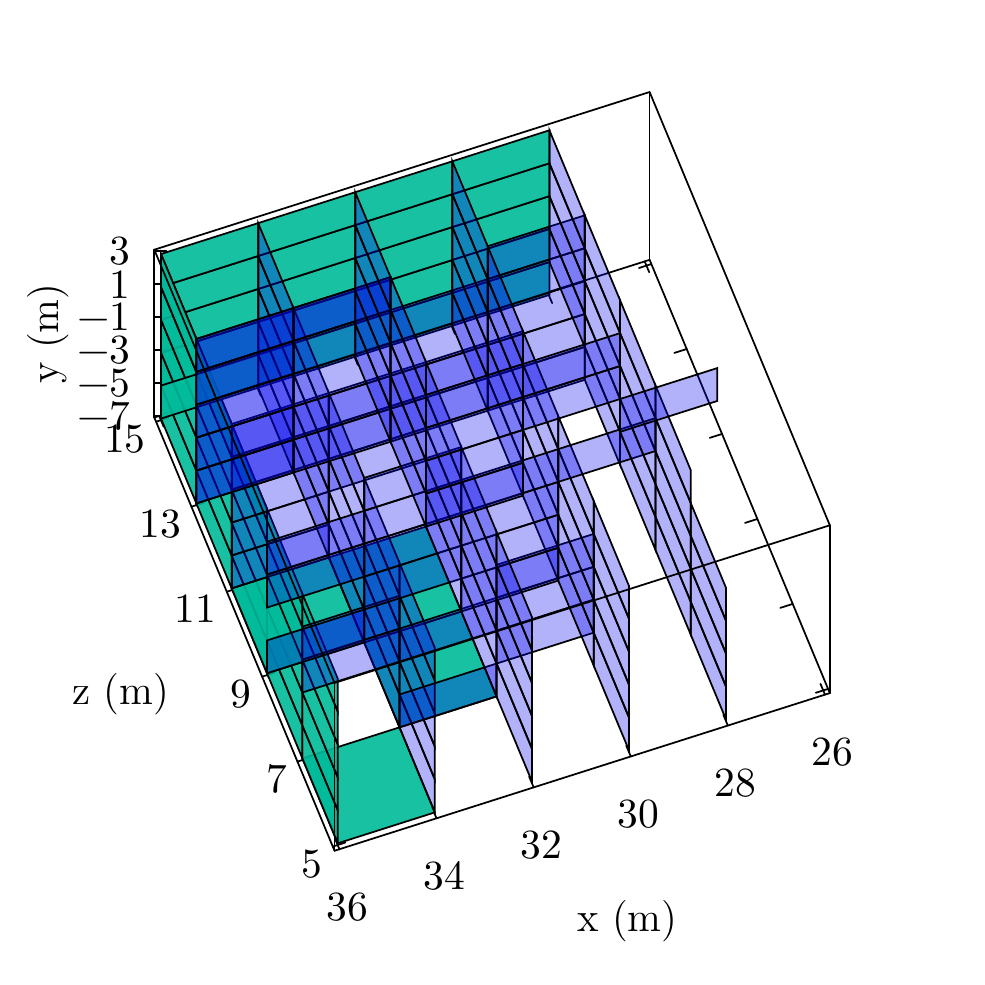}\hfil
	\caption{\textbf{Left:} Schematic representation of the benchmark-averaged hit weight ordered configuration, with $n = 150$ ($r_n = 0.42$). 
	\textbf{Right}: The same with $n = 250$ ($r_n = 0.71$).}
	\label{fig:nvc_geo}
\end{figure}

In Fig.~\ref{fig:nvc_geo} we show the configuration corresponding to the first 
$n = 150$ panels of this ordering ($r_n = 0.29$), which can achieve up to 
$80\%$ relative efficiency for the $h \to A' A'$ benchmarks, and $25$--$75$\% 
relative efficiencies for $b \to sS$. 
(This somewhat arbitrary choice for the number of panels 
combined with uncertainties from MC statistics 
leads to some planes being partially filled by disconnected panels.)
It is notable that:
(i) this configuration suggests predominant sensitivity is generated mainly by 
	the $z$-$y$ planes, 
	with some minimal contributions on the $z = 15$\,m face, and;
(ii) little to no role appears to be played by the external faces other than 
$x = 36$\,m. 
The selection of the $z$-$y$ internal faces comports with the expectation above that this estimator 
is correlated with minimizing distance of the first hit tracking layer from the LLP decay vertex.
We also show the configuration with the first $n = 250$ panels of 
this ordering ($r_n = 0.71$), which can achieve up to $90\%$ relative 
efficiency for the $h \to A' A'$ benchmarks, and $40$--$90$\% relative 
efficiencies for $b \to sS$. In this configuration, while the $z = 15$\,m face 
and internal $x$-$y$ faces now play a role, notably the top and bottom 
$y=-7$\,m and $y=3$\,m faces contribute little.

\begin{figure}[t]
	\includegraphics[width=0.4\linewidth]{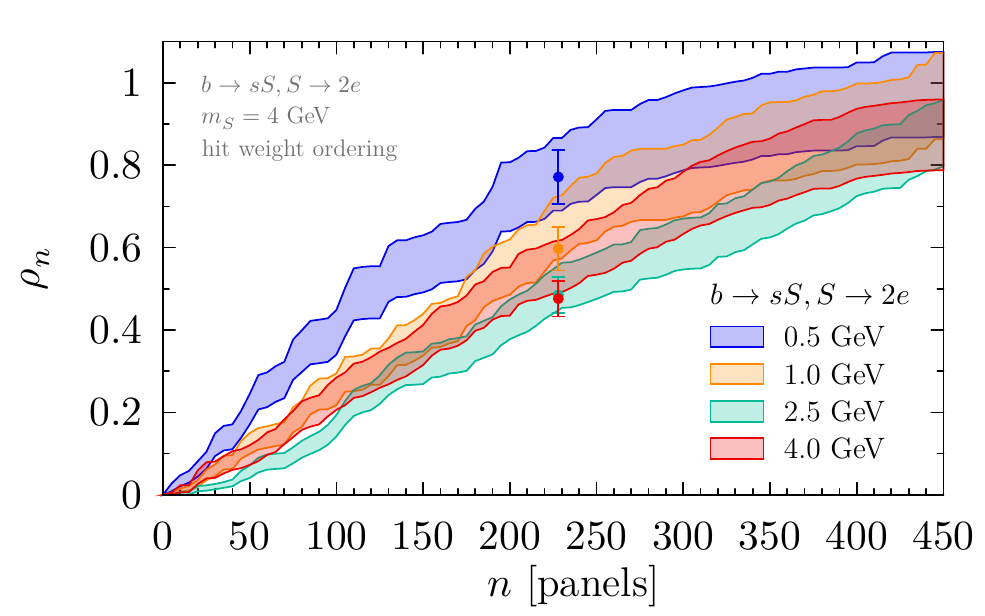}\hfil
	\includegraphics[width=0.4\linewidth]{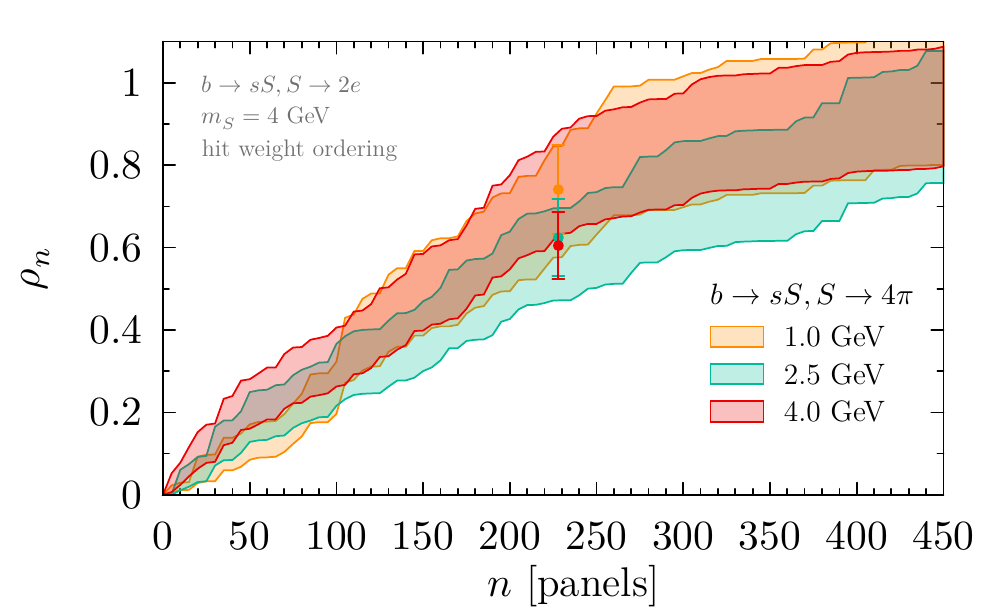}\\
	\includegraphics[width=0.4\linewidth]{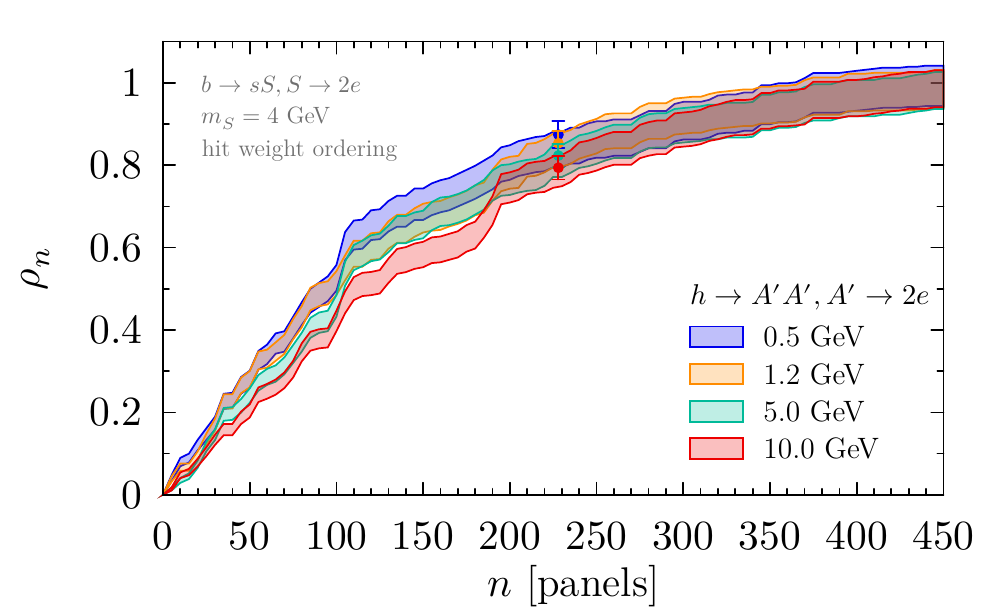}	
	\caption{The $1\sigma$ CL bands for relative vertex reconstruction efficiencies as a function of number of panels
	selected from the maximal envelope configuration,
	as determined by an ordering 
	using the hit weight from the $b \to sS$, $S \to 2e$ model with $m_S = 4$\,GeV.
	All uncertainties are from MC statistics.
	Also shown by correspondingly colored single data points are the relative vertex reconstruction efficiencies of the heuristic configuration of Fig.~\ref{fig:heur_geo}.}
	\label{fig:naiveordering4}
\end{figure}

In contrast to most of the benchmarks, in Fig.~\ref{fig:naiveordering} $\rho_n$
is almost linear in $n$ for the $b \to sS$, $S \to 2e$ model with 
$m_S = 4$\,GeV. To understand whether this can be improved, in 
Fig.~\ref{fig:naiveordering4} we show the relative reconstruction efficiencies 
using the ordering determined only by the hit weights for this $m_S = 4$\,GeV 
benchmark. While there is a mild improvement in the curvature of the 
$m_S = 4$\,GeV band, the $h \to A' A'$ curves become flatter, as do the
lighter $b \to sS$ benchmarks. Figures~\ref{fig:naiveordering} 
and~\ref{fig:naiveordering4} taken together suggest that there may exist 
configurations that can further optimize the efficiencies across the entire 
space of benchmarks.

\subsection{Optimized results}

In Appendix~\ref{app:opt} we present the formal construction of a branch-and-bound algorithm to solve this optimization problem.
It optimizes the increase in the objective function value per panel,
in order to generate an optimized partial ordering.
It makes use of a simple-to-compute bounding estimator of the objective 
function---the ``any-single-hit'' estimator, defined in Sec.~\ref{sec:anysnghit}.
As explained therein, this bounding estimator has the required algebraic properties to permit the computational complexity 
of this algorithm to scale approximately linearly in the number of panels $N$,
rather than the exponential scaling $\mathcal{O}(2^N)$ of a brute force approach.

\begin{figure}[htp!]
	\includegraphics[width = 0.55\linewidth]{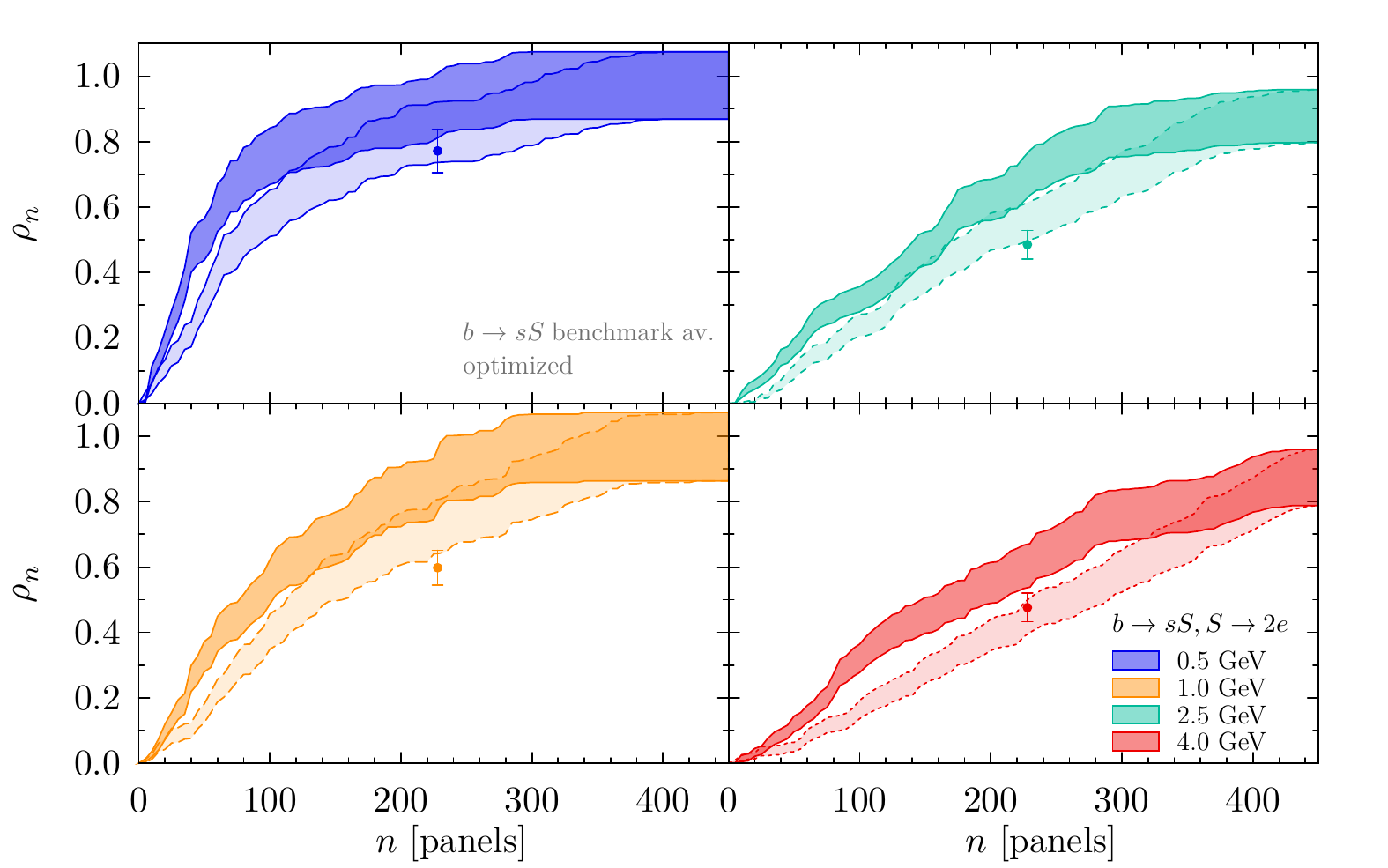}
	\includegraphics[width = 0.55\linewidth]{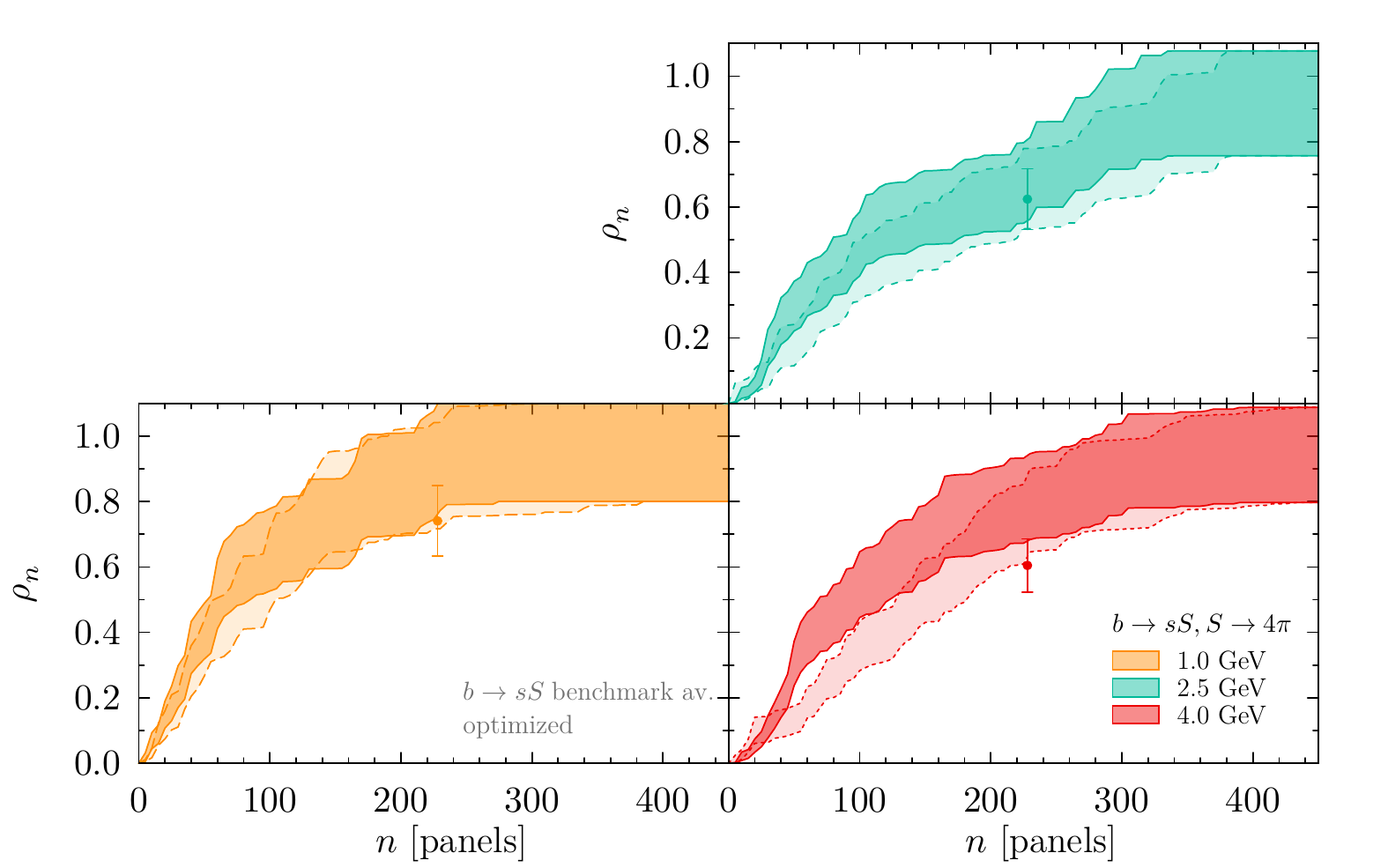}
	\includegraphics[width = 0.55\linewidth]{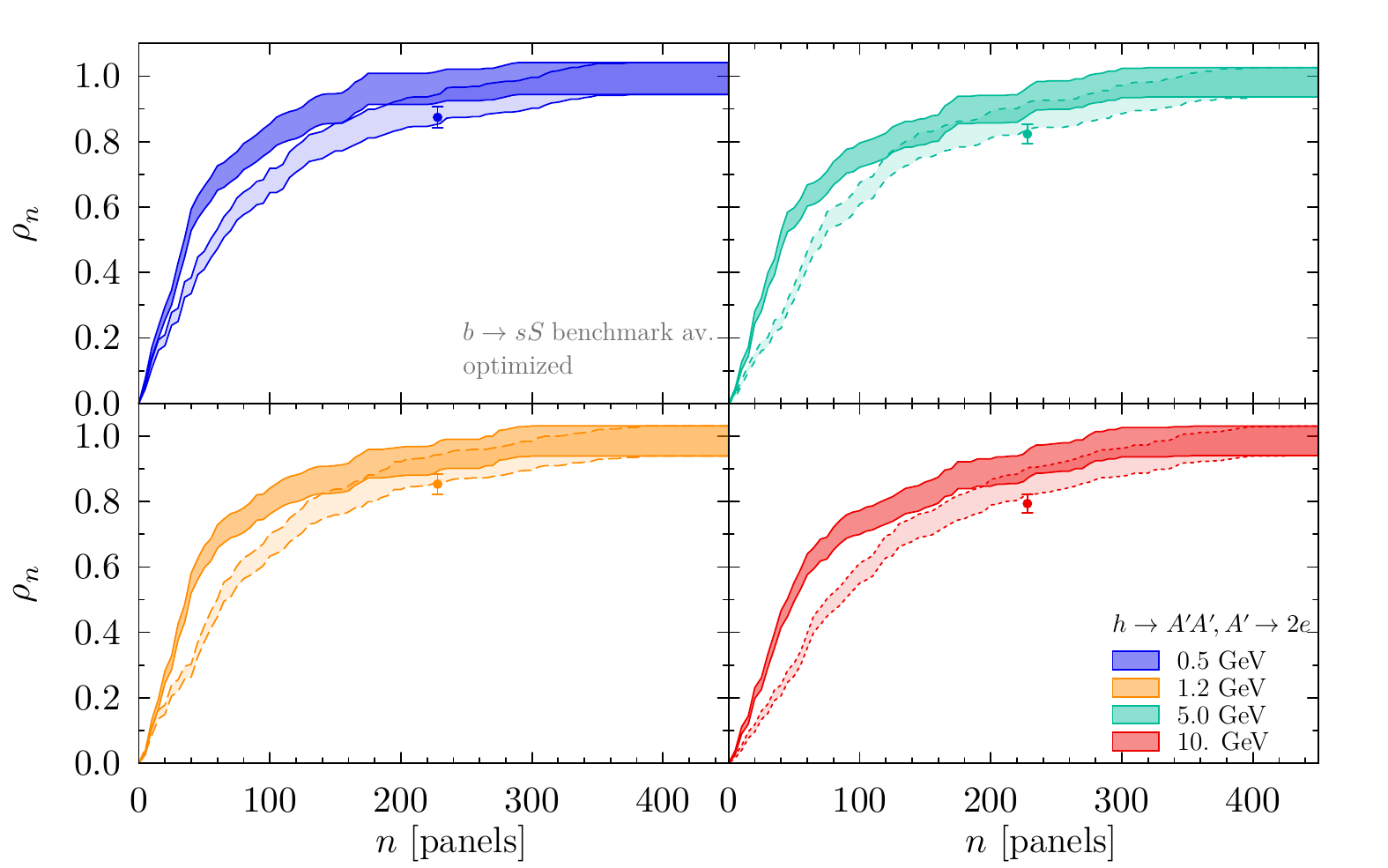}
	\caption{The $1\sigma$ CL bands for relative vertex reconstruction efficiencies (dark bands) as a function of number of panels
	selected from the maximal envelope configuration,
	as determined by the branch and bound optimization 
	applied to the event sample generated by the $b \to sS$, $S \to 2e$ model, 
	with objective function determined by a uniform prior over its four benchmark mass points.
	Also shown are the relative vertex reconstruction efficiencies generated by the benchmark-averaged hit weight as in Fig.~\ref{fig:naiveordering}
	(light bands).
	All uncertainties are from MC statistics.
	Also shown by correspondingly colored single data points are the relative vertex reconstruction efficiencies of the heuristic configuration of Fig.~\ref{fig:heur_geo}.}
	\label{fig:optordering}
\end{figure}

In Fig.~\ref{fig:optordering} we show the relative vertex reconstruction efficiencies 
for the partially optimized ordering generated by this branch-and-bound approach,
using the any-single-hit estimator applied to the objective determined by the $b \to sS$, $S \to 2e$ model.
This objective is averaged over the four mass benchmarks $m_S = 0.5$, $1.0$, $2.5$ and $4.0$\,GeV,
i.e. $\rho_n = \sum_{m_S = 0.5, 1.0, 2.5, 4.0\,\text{GeV}} \rho_{n,m_S}[B \to sS (2e)]$.
We find the partially optimized ordering contains $218$ groupings (see App.~\ref{app:optord}) of one or two panels for a total of $271$ panels 
selected from the maximal envelope configuration,
and one very large final grouping of $179$ panels, for which the optimizer does not specify a preferred ordering.
This behavior is not unexpected, as there are $50$ panels on the $x=25$\,m that receive zero hits, 
and many of the $75$ panels on the $z = 5$, $7$ and $9$\,m faces receive significantly fewer hits than others in the envelope configuration,
along with some on the $y = 3$ and $-7$\,m faces.
To further order this last grouping, we apply the naive benchmark-averaged hit weight estimator
on the space of events whose LLP vertices  are \emph{not} reconstructed by these first $271$ panels.

The subsequent result for each of the eleven benchmarks, shown by dark bands, 
is compared to the corresponding relative vertex reconstruction efficiencies generated 
by the naive benchmark-averaged hit weight estimator in Fig.~\ref{fig:naiveordering},
shown by light bands.
For all eight of the $b \to sS$, $S \to 2e$ and $h \to A'A'$, $A' \to 2e$  benchmarks,
the optimized orderings outperform the naive hit weight estimator, 
by up to $\sim 20\%$ in the small $n$ regime.
We see that for e.g. 150 panels ($r_n = 0.29$), 
the relative vertex reconstruction efficiency $\rho_n$ is $60$--$80$\% for $b \to sS$ portals, 
and $80$--$90$\% for $h \to A'A'$.
The optimized orderings also achieve a notably higher efficiency than the heuristic configuration of Fig.~\ref{fig:heur_geo}.
The increase in performance for the $h \to A'A'$ portals occurs even though the tracks for these benchmarks are typically harder and more collinear
than the $b \to sS$, $S \to 2e$ benchmarks on which the optimizer objective was defined.
This is not unexpected since the $b \to sS$, $S \to 2e$ benchmarks would demand a more widely distributed configuration of panels to achieve significant sensitivity, 
and thereby capture the more boosted $A'$ decays.
For the three $b \to sS$, $S \to 4\pi$ benchmarks, by contrast, the performance between naive and optimized approaches is similar:
the distribution of tracks for these benchmarks are broader and softer compared to the $b \to sS$, $S \to 2e$ benchmarks, 
so that further gains in sensitivity would likely require prioritizing an even more widely distributed configuration of panels.

\begin{figure}[bt]
      	\includegraphics[width=7cm]{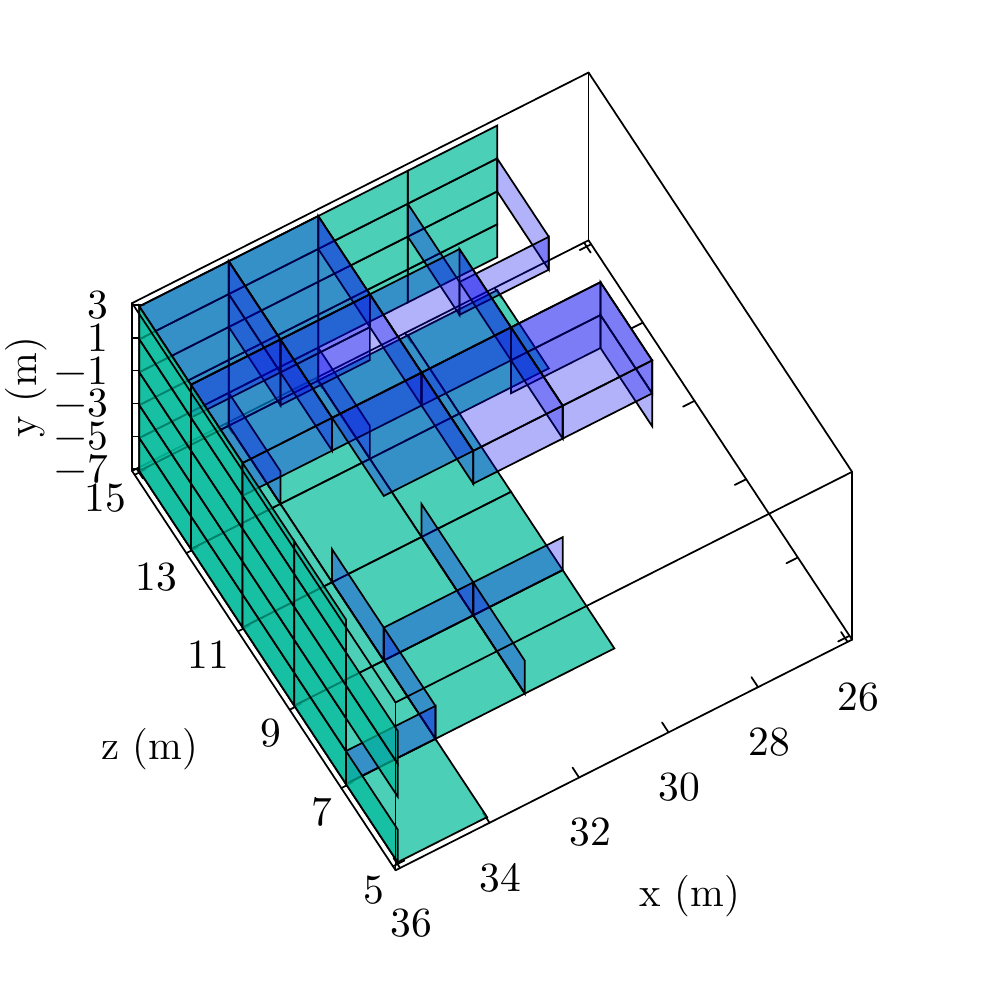}\hfil
	\includegraphics[width=7cm]{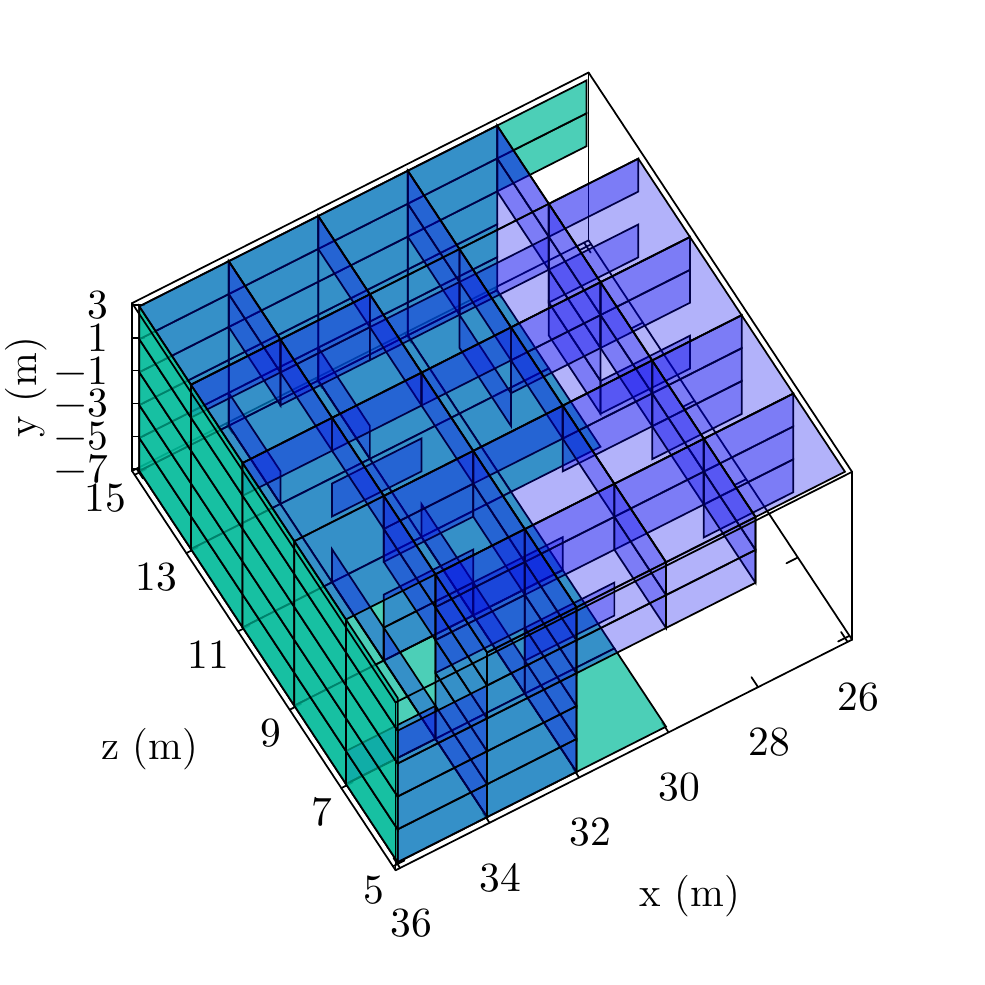}\hfil
	\caption{\textbf{Left:} Schematic representation of the benchmark-averaged branch and bound optimized configuration, with $n = 150$ ($r_n = 0.42$). 
	\textbf{Right}: The same with $n = 250$ ($r_n = 0.71$). }
	\label{fig:opt_geo}
\end{figure}

In Fig.~\ref{fig:opt_geo} we show the panel configurations for $n = 150$ and $250$ panels, 
corresponding to the partially optimized ordering generated by the branch-and-bound approach.
Intrinsic uncertainties from the MC statistics lead to some ``noise'' in the selected panels.
With fewer panels, the optimizer favors mainly panels on the external faces, 
as expected from the requirement that each track hits at least two RPC triplets.
Where double RPC triplets on an external face are not available, such as on the $y=3$\,m face, 
the optimizer selects instead panels on the adjacent edge of the internal faces.
As more panels are added, the optimizer further fills out these adjacent edges on the internal faces,
leading to a partial ring-like configuration on the internal faces.

\section{Summary}
In this work we demonstrated that compared to prior baseline hermetic RPC configurations for \CODEXb-like detectors, 
a significant optimization of the amount of required RPC-instrumented tracking surface is achievable.
We showed that nested optimal detector configurations, 
which optimize the LLP decay vertex reconstruction efficiency as a function of the number of tracking station 
elements---here, the number of RPC triplet panels,
can be efficiently computed using an ``any-single-hit'' bounding estimator combined with a branch-and-bound optimization method.
The computation time is approximately linear in the number of panels, rather than exponential.
This is achievable because of the set-theoretic properties of the bounding estimator, 
that permit identification of a directed path(s) through the powerset configuration of panels
on which the objective function---the vertex reconstruction efficiency---is maximized and nondecreasing.

Within this approach, we considered the vertex reconstruction efficiency for a range of different LLP production and decay portals,
chosen to provide good coverage over the space of typical LLP production and decay morphologies and kinematics:
In particular the $b$-hadron production portals $b \to sS$, $S \to 2e$ and  $b \to sS$, $S \to 4\pi$, 
and the Higgs production portal $h \to A'A'$, $A'\to 2e$, for scalar LLP $S$ and vector LLP $A'$, respectively, 
over a range of different LLP masses.
For e.g. 150 panels ($r_n = 0.29$), the optimized configurations achieve
the relative vertex reconstruction efficiencies of $60$--$80$\% for $b \to sS$ portals, 
and $80$--$90$\% for $h \to A'A'$ compared to the baseline configuration, which has $400$ panels.

The branch-and-bound optimized configurations attained higher relative efficiencies compared to a more naive optimization approach, 
that ordered the panels based simply on the LLP track hit weight density.
The latter is, however, even faster to compute.
We found it typically provides a good proxy for characterizing 
the degree of optimization that is possible in any given LLP model,
though it preferred a notably different configuration of tracking layers, in particular by selecting panels on internal faces earlier in the ordering.
(It thus appears correlated with minimization of the distance of first-hit tracking layer from the LLP decay vertex.)
However, the branch-and-bound optimizer is capable of deducing configurations that better avoid redundancies in hit-hit correlations among different panels,
and thus should be preferably used for optimization of LLP detectors.
Although we used a simplified set of track vertex reconstruction criteria in this study,
future applications of this algorithm 
may straightforwardly incorporate more realistic requirements for track reconstruction and LLP vertex resolution.
One may thus implement a realistic optimization analysis for a future technical design report
that incorporates buildability and other engineering constraints.

Underpinning this analysis was the development of a new, generalized simulation framework for the computation of LLP vertex reconstruction efficiencies
across a wide range of LLP portals and different detector geometries.
Beyond enabling the optimization of detector geometries for LLP detection, 
the combined simulation and optimization framework may also be adapted to study background rejection,
in order to optimize the amount of require passive and active shielding in a realized LLP detector.
This will be a subject of future work.

\acknowledgements
We thank the membership of the CODEX-b collaboration for discussions, 
and in particular Juliette Alimena, Xabier Cid Vidal, Vladimir Gligorov, Phillip Ilten, Daniel Johnson, Titus Momb\"acher,  Emilio Rodr\'\i guez Fern\'andez, and Michele Papucci 
for comments on the manuscript.
SK thanks the Aspen Center for Physics (supported by the NSF Grant PHY-1607611) for its hospitality, where part of this work was performed.
This work was supported by the Laboratory Directed Research and Development Program of Lawrence Berkeley National Laboratory 
under U.S. Department of Energy Contract No. DE-AC02-05CH11231.
SK, BN and DJR are supported by the Office of High Energy Physics of the U.S. Department of Energy under Contract No. DE-AC02-05CH11231. 

\appendix

\section{Optimization algorithm}
\label{app:opt}
\subsection{Optimized partial orderings}
\label{app:optord}
In principle, one needs `merely' to compute the objective function over the power set of $\Sigma$, 
in order to determine $\sigmaopt{n}$ for any given $n$.
For the case that $N$ is of the order of several hundred panels, 
however, this becomes exponentially hard and computationally prohibitive.
Further, because of the nesting requirement~\eqref{eq:chainofopt}, 
it is not sufficient to identify the globally optimal subset of any particular cardinality. 
Rather, we seek to identify a sequence of $\sigmaopt{n}$ such that $f(\sigmaopt{n})$ 
has the maximal possible negative curvature in $n$.

To understand the character of this optimization problem,
suppose the entire set of panels can be partitioned 
into a set of $K$ disjoint subsets---``groupings''---of panels, 
\mbox{$\Sigma = \dsigma_1\cup\ldots\cup\dsigma_{K}$}.
We use the notation for the $i$th grouping ``$\dsigma_i$'' to emphasize that a grouping  is a (typically small) set of panels to be added to a larger configuration.
We further define $\Sigma^k = \dsigma_{1} \cup\ldots\cup\dsigma_{k}$ for each $k \le K$, with $\Sigma^0 = \varnothing$, the empty set.
With respect to the conditional objective function
\begin{equation}
	f(\dsigma | \sigma) \equiv f(\dsigma\cup\sigma) - f(\sigma)\,,
\end{equation}
which describes the contribution to the objective from adding $\dsigma$ to a given configuration of panels $\sigma$,
the total objective may always be expanded as a series of positive semidefinite partial sums
\begin{align}
	f(\Sigma) 
	&= f(\dsigma_{1}) + f(\dsigma_{2} | \dsigma_{1}) + f(\dsigma_{3}| \dsigma_{1} \cup \dsigma_{2}) + \ldots + f(\dsigma_{K} | \dsigma_1\cup\ldots\cup\dsigma_{K-1}) \nn\\
	& = f(\dsigma_1|\Sigma^0) + f(\dsigma_2| \Sigma^1) + f(\dsigma_3| \Sigma^2) + \ldots + f(\dsigma_K|\Sigma^{K-1})\,. \label{eqn:objectiveseriesexp}
\end{align}
If it is the case that the partition has the property
\begin{equation}
	\label{eqn:localcond}
	f(\dsigma_i |\Sigma^k) \ge f(\dsigma_i |\Sigma^{k+1})
\end{equation}
for all $i$, and $\dsigma_{k+1}$ satisfies 
\begin{equation}
	\label{eqn:maxcond}
	f(\dsigma_{k+1}|\Sigma^k) = \max_i f(\dsigma_i |\Sigma^k)\,,
\end{equation}
then each partial sum is bounded above by the previous one,
because $f(\dsigma_k|\Sigma^{k-1}) = \max_i f(\dsigma_i |\Sigma^{k-1}) \ge f(\dsigma_{k+1} |\Sigma^{k-1}) \ge f(\dsigma_{k+1} |\Sigma^{k})$.
The maximization ensures that $f(\Sigma^k)$ has maximal negative semidefinite curvature over $n$,
in the strict sense that one has maximized each term only on the set of the $\{\dsigma_i\}$ in the partition.
Thus one may identify $\Sigma^k = \sigmaopt{n_k}$ in the case that Eqs.~\eqref{eqn:localcond} and~\eqref{eqn:maxcond} hold.
Note, however, for a given $\Sigma^k$ possible degenerate choices may arise for $\dsigma_{k+1}$,
so that multiple optimized orderings may exist: the solution to the optimization problem is not unique. 

The correspondence in Eq.~\eqref{eqn:objectiveseriesexp} makes it clear that we can consider $\dsigma_k$ 
satisfying Eqs.~\eqref{eqn:localcond} and~\eqref{eqn:maxcond}
as defining the step between $\sigmaopt{n_{k-1}}$ and $\sigmaopt{n_{k}}$.
In this sense, let us represent the elements of the power set $\sigma\in\mathcal{P}(\Sigma)$ 
as the vertices of a graph connected by edges,
which each represent the addition or subtraction of a single panel between two adjacent vertices (see Fig. \ref{fig:hypercube}).
We consider only the two \emph{directed} graphs, in which the edges in a path between two vertices are all additive or are all subtractive
(since we seek an ordering that represents a prioritization of the panels, not a set of rearrangements).
We refer to these as the ``upwards'' and ``downwards'' graph, respectively.
Each vertex is weighted by the objective $f$.
Each panel-wise expansion of the objective
\begin{equation}
	\label{eq:panelexpansion}
	f(\Sigma) = f(\{p_1\}) + f(\{p_2\}|\{p_1\}) + f(\{p_3\}|\{p_1,p_2\}) + \ldots + f(\{p_n\}|\Sigma\setminus\{p_n\})\\,,
\end{equation}
is specified by a path through the directed graph, that includes panels in the order they appear in the sequence $\{p_i\}$.
Each term of Eq.~\eqref{eq:panelexpansion} encodes the incremental changes in the objective along this path. 
A grouping $\dsigma_k = \{p_{n_{k-1}+1}, \ldots, p_{n_{k}}\}$ is the subpath between the vertex  $\sigmaopt{n_{k-1}}$ and vertex $\sigmaopt{n_{k}}$.

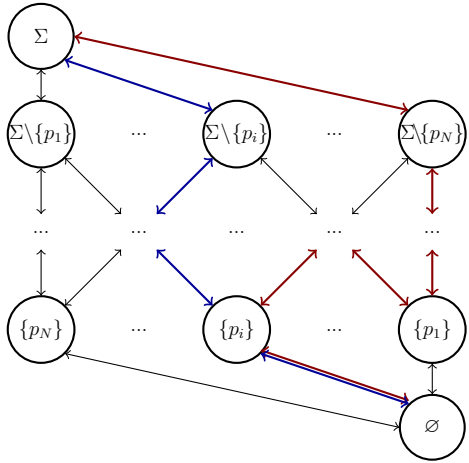
\begin{figure}[h]
\label{fig:hypercube}
\centering
\begin{tikzpicture}[node distance=2cm, 
	setnode/.style={circle, draw=black!100, thick, inner sep = 0pt, text width = 1.3cm, align = center}, 
	dotnode/.style={rectangle, draw=none, thick, minimum size = 7mm}, 
	scale = 0.65, transform shape]
\node[setnode]        (all)                             {$\Sigma$};
\node[setnode]        (Ama1)       [below of= all]      {$\Sigma\!\setminus\! \{p_1\}$};
\draw[<->] (all.south) -- (Ama1.north); 
\node[dotnode] (layer1dotsleft)   [right of=Ama1]       {$...$};
\node[setnode]        (Amai)       [right of= layer1dotsleft]     {$\Sigma\!\setminus\! \{p_i\}$};
\draw[<->, thick, draw=blue!60!black] (all.south east) -- (Amai.north west); 
\node[dotnode] (layer1dotsright)   [right of=Amai]       {$...$};
\node[setnode]        (AmaM)       [right of= layer1dotsright]     {$\Sigma\!\setminus\! \{p_N\}$};
\draw[<->, thick, draw=red!60!black] (all.east) -- (AmaM.north west); 
\node[dotnode]        (dots1)      [below of= Ama1]      {$...$};
\draw[<->] (Ama1.south) -- (dots1.north);
\node[dotnode]        (dots2)      [right of= dots1]      {$...$};
\draw[<->] (Ama1.south east) -- (dots2.north west);
\draw[<->, thick, draw=blue!60!black] (Amai.south west) -- (dots2.north east);
\node[dotnode]        (dots3)      [right of= dots2]      {$...$};
\node[dotnode]        (dots4)      [right of= dots3]      {$...$};
\draw[<->] (Amai.south east) -- (dots4.north west);
\draw[<->] (AmaM.south west) -- (dots4.north east);
\node[dotnode]        (dots5)      [right of= dots4]      {$...$};
\draw[<->, thick, draw=red!60!black] (AmaM.south) -- (dots5.north);
\node[setnode]        (aM)         [below of= dots1]     {$\{p_N\}$};
\draw[<->] (dots1.south) -- (aM.north);
\draw[<->] (dots2.south west) -- (aM.north east);
\node[dotnode] (layer3dotsleft)    [right of=aM]        {$...$};
\node[setnode]        (ai)         [right of=layer3dotsleft]  {$\{p_i\}$};
\draw[<->, thick, draw=blue!60!black] (dots2.south east) -- (ai.north west);
\draw[<->, thick, draw=red!60!black] (dots4.south west) -- (ai.north east);
\node[dotnode] (layer3dotsright)    [right of=ai]        {$...$};
\node[setnode] (a1)    [right of=layer3dotsright]        {$\{p_1\}$};
\draw[<->, thick, draw=red!60!black] (dots5.south) -- (a1.north);
\draw[<->, thick, draw=red!60!black] (dots4.south east) -- (a1.north west);
\node[setnode]       (empty)     [below of=a1]           {$\varnothing$};
\draw[<->] (a1.south) -- (empty.north);
\draw[<->, thick, draw=red!60!black] ([shift=({0,2pt})]ai.south east) -- ([shift=({0,2pt})]empty.north west);
\draw[<->, thick, draw=blue!60!black] (ai.south east) -- (empty.north west);
\draw[<->] (aM.south east) -- (empty.west);
\end{tikzpicture}
\caption{A simplified drawing of the graph of the $2^N$ elements in $\mathcal{P}(\Sigma)$.
	Each arrow between vertices defines an edge of the graph that adds or subtracts an individual panel.
	The drawn vertices resemble the structure of the power set of $3$ elements with all others 
	implied by ellipses. 
	The blue path consists of strictly additive or subtractive edges, and is an allowed path in the upwards or downwards directed graph;
	the red path is neither strictly additive nor subtractive, and not an allowed path within a directed graph.
	Identifying $\sigmaopt{n}$ amounts to finding the vertex in the $n$th layer with the largest objective.} 
\label{fig:hypercube}
\end{figure}

In the graph picture, 
the optimization problem can now be constructed on the space of possible paths through the directed graph, 
rather than in terms of the choice of subsets of $\Sigma$.
In the directed graph, there are $2^N$ possible paths, 
so that it remains equally impractical to consider all possibilities iteratively as to compute $f$ on all vertices.
Instead, we shall seek a method of recursively constructing a path through the graph,
that identifies $\dsigma_k$ satisfying the conditions~\eqref{eqn:localcond} and~\eqref{eqn:maxcond},
and hence the optimized partial ordering $\sigmaopt{n_k}$.

To begin, we observe that $\sigmaopt{n_1}$ and $\sigmaopt{n_{K-1}}$ are
(relatively) easy to construct, up to degeneracy, using the following two methods, respectively:
\begin{enumerate}[wide, labelwidth=!, labelindent=2em, label = \textbf{(\Roman*)}, noitemsep, topsep =0pt]
\item Starting from the empty configuration $\varnothing = \sigmaopt{0}$,
	progress through the upwards directed graph computing the vertex weights,
	and identify the set of vertices of the smallest cardinality with $f(\sigma) > f(\varnothing) (=0)$. 
	Of these, choose the vertex with the largest objective weight to be $\sigmaopt{n_1}$.	
	For the LLP decay vertex reconstruction requirements considered here (see Sec.~\ref{sec:recoreq}) 
	this choice will often be just two panels, in which case $n_1 = 2$. 
	However, an objective with more complex reconstruction requirements, 
	such as more than two tracks per LLP decay vertex, might select $n_1 >2$.
	By construction, all $\sigma$ such that $|\sigma| \leq n_1$ must satisfy
	\begin{equation}
		f(\sigmaopt{n_1}) \geq f(\sigma)\,.
	\end{equation}
	\label{it:up}
\item Starting from the full configuration $\Sigma = \sigmaopt{n_K}$ (under our convention $n_K = N$)
 	and moving through the downwards directed graph,
 	find the configuration of smallest cardinality whose objective is the same as the total $f(\sigma) = f(\sigmaopt{n_K})$,
	and choose this to be $\sigmaopt{n_{K-1}}$.
	Any vertex $\sigma$ on the path from $\Sigma$ to this configuration must satisfy, 
		  \begin{equation}
		  	\label{eqn:effperpanel}
			f(\sigmaopt{n_{K-1}})/n_{K-1} \geq f(\sigma)/|\sigma|\,,
		  \end{equation}
	as must all $\sigma$ such that $|\sigma| \geq n_{K-1}$. 
	That is, $\sigmaopt{n_{K-1}}$ has the maximal objective per panel among all configurations of size $n_{K-1}$ or higher.
	\label{it:down}
\end{enumerate}

It is important to remember that neither $\sigmaopt{n_1}$ nor $\sigmaopt{n_{K-1}}$ constructed in this manner are guaranteed to be unique,
though one must require $\sigmaopt{n_{K-1}} \supset \sigmaopt{n_1}$.
Iteratively applying the algorithms \ref{it:up} and \ref{it:down} starting with $\sigmaopt{n_1}$ and $\sigmaopt{n_{K-1}}$,
however, does not identify $\sigmaopt{n_2}$ and $\sigmaopt{n_{K-2}}$.
The reason for this is that although a directed subgraph from $\sigmaopt{n_1}$ to $\sigmaopt{n_{K-1}}$ has the same topology as the full graph 
(they are hypercubes of dimension $n_{K-1} - n_1$ and $N$, respectively)
one cannot guarantee from \ref{it:up} and \ref{it:down} alone that $\sigmaopt{n_2}$ and $\sigmaopt{n_{K-2}}$ lie within this subgraph.
Instead, we seek a method to identify a \emph{subset} of the $\sigmaopt{n_k}$ that is sparsely distributed over the entire graph,
which we denote as $\sigmaopt{n_{k_i}}$.
This may then be applied recursively to generate a larger and more refined set of $\sigmaopt{n_k}$ and hence an optimized partial ordering.
The advantage of this method will be that it scales computationally as $\mathcal{O}(K 2^J)$, 
where $J$ is the typical step size of adjacent $n_{k_i} - n_{k_{i+1}}\sim \text{few}$,
rather than as $\mathcal{O}(2^N)$.

\subsection{Branch and bound algorithm}
\label{sec:branchbound}
To proceed further, we seek a means to traverse the upwards directed graph from a given $\sigmaopt{n_k}$, 
and efficiently identify which of its child vertices cannot themselves have children that may be a candidate $\sigmaopt{n_{k'}}$, $n_{k'} > n_k$.
This allows us to prune away paths, leaving a small subset on which we evaluate the objective directly to identify $\sigmaopt{n_{k'}}$.
To do this, we note from Eq.~\eqref{eqn:effperpanel}, that $f(\sigmaopt{n_k})/n_k$ should similarly be maximal for all $|\sigma| = n_k$.
Thus we expect that a candidate $\sigmaopt{n_{k'}}$ should lie on the directed path on which the \emph{objective per panel} is maximal.
This leads us to propose the following test function for a vertex $\sigma \supset \sigmaopt{n_k}$,
namely the maximal objective per panel of the children of $\sigma$,
\begin{equation}
	\mathcal{T}(\sigma) = \max_{\sigma' \subset \sigma}\frac{f(\sigma')}{|\sigma'|} = \max_{\dsigma: \dsigma \cap \sigma = \varnothing}\frac{f(\sigma) + f(\dsigma|\sigma)}{|\dsigma| + |\sigma|}\,.
\end{equation}
If $\mathcal{T}(\sigma) < f(\bar\sigma)/|\bar\sigma|$ for some other $\bar\sigma \supseteq \Sigma^k$
for which the objective is known,
then all paths from $\sigma$ upwards can be pruned.

In practice $\mathcal{T}$ is too expensive to compute, as it requires computation of $2^{N-|\sigma|}$ terms.
We therefore require a bounding estimator of the objective, $\fbar(\sigma) > f(\sigma)$,
from which we can construct a bounding estimator of $\mathcal{T}$ that can be quickly computed with only linear complexity in $N-|\sigma|$.
In particular, provided $\fbar$ has the conditional subadditive property that
\begin{equation}
	\label{eqn:subadd}
	\fbar(\dsigma \cup \dsigma'| \sigma) \le \fbar(\dsigma|\sigma) + \fbar(\dsigma'|\sigma)\,,
\end{equation}
then
\begin{equation}
	\label{eqn:boundobj}
	\mathcal{B}(\sigma) = \max_{\dsigma: \dsigma \cap \sigma = \varnothing} \frac{f(\sigma) + \sum_{p \in \dsigma} \fbar(\{p\}| \sigma)}{|\dsigma| + |\sigma|} \ge \mathcal{T}(\sigma)\,.
\end{equation}
That is, $\mathcal{B}(\sigma)$ bounds the test function, but requires only the computation of $N - |\sigma|$ terms,
and is therefore far faster to compute.
It is then sufficient to test whether $\mathcal{B}(\sigma) < f(\bar\sigma)/|\bar\sigma|$ for $\bar\sigma \supseteq \sigma$ whose objective is known, 
in order to eliminate all paths upwards from $\sigma$.

The algorithm to construct $\sigmaopt{n_{k'}}$ from $\sigmaopt{n_k}$ then proceeds as follows. 
Define a set of vertices $\Sigma_{\text{eval}}$, initially containing just $\sigmaopt{n_k}$, whose objective $f(\sigmaopt{n_k})$ is known.
Starting from $\sigmaopt{n_k}$:
\begin{enumerate}[label = \textbf{(\arabic*)}, noitemsep, topsep =0pt]
\item  Compute the objective for the immediate child vertices,
i.e., $f(\sigma)$, for each $\sigma = \{p\} \cup \sigmaopt{n_k}$ with $p \not\in \sigmaopt{n_k}$.
Assign each $\sigma$ to $\Sigma_{\text{eval}}$.
\label{bb:obj}
\item For each such $\sigma$:
	\begin{enumerate}  [label = \textbf{(\roman*)}, noitemsep, topsep =0pt]
 	\item Compute $\fbar(\{p\}| \sigma)$ for all $p \not\in \sigma$ and sort them from highest to lowest value, to generate the ordered sequence $\{\fbar(\{p_i\}| \sigma)\}_i$.
	\item One may then compute the maximand in Eq.~\eqref{eqn:boundobj}
	\begin{equation}
		\frac{f(\sigma) + \sum_{i=1}^n\fbar(\{p_i\}| \sigma)}{n + |\sigma|}\,,
	\end{equation}
 	as a function of $n$ until a local maximum is reached.
	\item This local maximum is used as a proxy for the true maximum, and thus the value of $\mathcal{B}(\sigma)$.
	If $\mathcal{B}(\sigma) < f(\sigma')/|\sigma'|$ for some other $\sigma' \in \Sigma_{\text{eval}}$ whose objective has been computed in step~\ref{bb:obj}, 
	then discard $\sigma$ and all paths in the directed graph that pass through it, and remove $\sigma$ from $\Sigma_{\text{eval}}$.
	\end{enumerate}
\item For each remaining child vertex $\sigma$, recursively apply this algorithm starting from step~\ref{bb:obj}. 
	Each recursion branch terminates when all immediate children are discarded by the bound.
\item Once all branches terminate, the $\sigma \in \Sigma_{\text{eval}}$ with the largest $f(\sigma)/|\sigma|$ is identified as $\sigmaopt{n_{k'}}$.
\end{enumerate}

Having determined an initial sparse set, $\sigmaopt{n_{k_i}}$ 
one may then further recursively apply this entire algorithm to each restricted directed subgraph from $\sigmaopt{n_{k_i}}$ to $\sigmaopt{n_{k_{i+1}}}$---a 
hypercube of dimension $n_{k_{i+1}} - n_{k_i}$ that is self-similar to the full graph---in
order to determine a finer-spaced set of optimized configurations.
Once restricted to this subgraph, the possible members of $\Sigma_{\text{eval}}$ at each step is correspondingly more restricted,
leading to less restrictive bounds.
Thus one may identify optimized configurations excluded in a prior pass.
This recursion terminates once no further optimized configurations are identified, typically when $n_{k+1} - n_k \sim \text{few}$.
This procedure generates the partially optimized ordering $\sigmaopt{n_k}$, for a set of $n_k \le N$.

\subsection{Any-single-hit estimator}
\label{sec:anysnghit}
We now seek to identify a bounding estimator that obeys Eq.~\eqref{eqn:subadd}.
For the kinds of objectives and reconstruction criteria we consider, 
the objective function $f$ may be decomposed into two pieces. 
First, a ``reconstruction map'', that determines the set of events that is reconstructible by a configuration, $g : \sigma \mapsto g(\sigma)$.
This map must be positive semi-definite, i.e. $g(\sigma) \subseteq g(\sigma')$ for $\sigma \subseteq \sigma'$.
Second, an outer measure $\rho$ that determines the vertex reconstruction efficiency of $\sigma$,
i.e. $\rho: g(\sigma) \mapsto f(\sigma)$.
In the composition $f = \rho \circ g$, $\rho$ is additive (strictly, countably additive) by construction,
and independent of reconstruction criteria. 

\begin{figure}[t]
\centering
\includegraphics[width=0.45\linewidth]{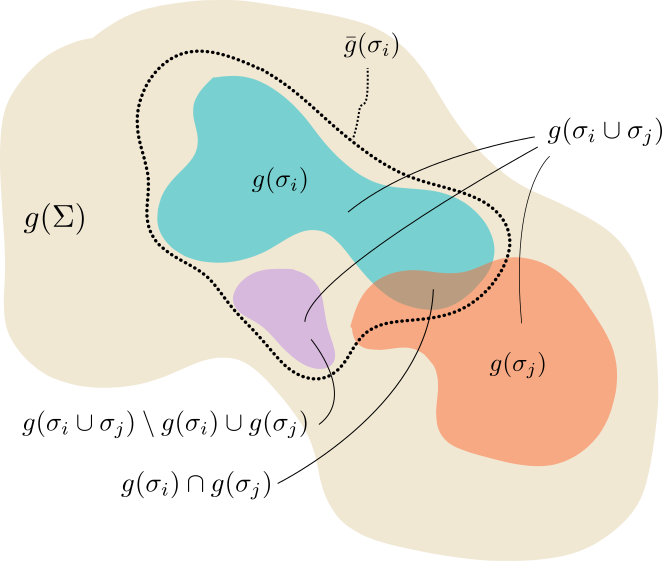}
\includegraphics[width=0.45\linewidth]{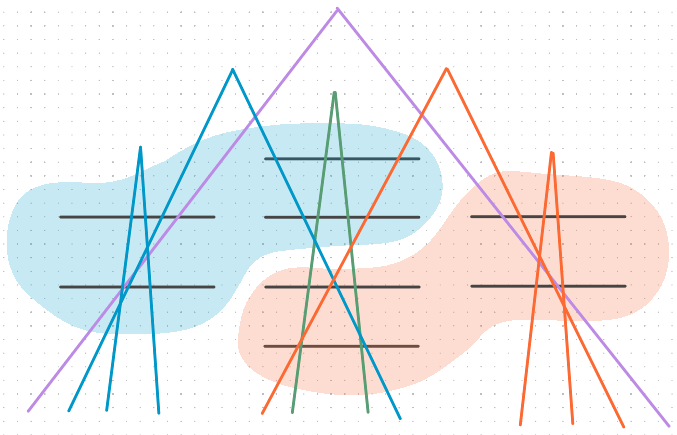}
\caption{Left: An abstract picture in the space of LLP events. 
Right: Sample event track topologies, with each event colored corresponding to which panel grouping (red or blue) that reconstructs it.
The green event is reconstructed by either 
the red grouping or blue grouping alone (it is in $g(\sigma_\text{red})\cap g(\sigma_\text{blue})$) in contrast to the purple event which 
needs both groupings to satisfy a 2-hit criteria on all its tracks 
(it is in $g(\sigma_\text{red}\cup\sigma_\text{blue}) \setminus g(\sigma_\text{red})\cup g(\sigma_\text{blue})$). 
The dotted region $\bar{g}(\sigma_\text{blue})$ denotes events selected by a looser ``bounding" criteria than $g$, 
that is constructed so that $\bar{g}(\sigma\cup\sigma') = \bar{g}(\sigma)\cup\bar{g}(\sigma')$ for all $\sigma, \sigma'\subseteq\Sigma$.}
\label{fig:eventspaceblobs}
\end{figure}

In this language, the negative nonadditivity ($f(\sigma \cup \sigma') < f(\sigma) + f(\sigma')$) 
of the objective from panel-panel redundancies 
arises because for two disjoint configurations $\sigma$ and $\sigma'$, the intersection
$g(\sigma) \cap g(\sigma')$ can nonetheless be non-empty.
Similarly, the positive nonadditivity  ($f(\sigma \cup \sigma') >  f(\sigma) + f(\sigma')$) 
from hit-hit correlations arises because the set of events in $g(\sigma\cup\sigma')$
but not in $g(\sigma)\cup g(\sigma')$---i.e.
$g(\sigma\cup\sigma') \setminus \big[ g(\sigma)\cup g(\sigma')\big]$---can be nonempty.
See Figure \ref{fig:eventspaceblobs}.

The LLP vertex reconstruction map $g$
requires at least two LLP decay tracks to hit at least two RPC triplets, 
with a minimum track momentum threshold and hit separation (see Sec.~\ref{sec:recoreq}).
These requirements can be subsumed by looser ones:
in particular the loosest, nontrivial such requirement is simply for
at least one track in an event to have at least one hit.
We define the corresponding ``any-single-hit'' reconstruction map by $\bar{g}$.
Though $\bar{g}$ does not permit actual reconstruction of the LLP decay vertex,
it must be the case that
\begin{equation}
	\bar{g}(\sigma) \supseteq g(\sigma)\,,
\end{equation}
for any $\sigma$. Thus defining the any-single-hit estimator $\fbar = \rho \circ \bar{g}$, then 
\begin{equation}
	f(\sigma) = \rho(g(\sigma)) \leq \rho(\bar{g}(\sigma)) = \fbar(\sigma)\,,
\end{equation}
i.e. $\fbar$ bounds $f$.
This estimator is clearly positive semi-definite. Moreover, $\bar{g}$ is additive by construction,
\begin{equation}
	\bar{g}(\sigma\cup\sigma') = \bar{g}(\sigma)\cup\bar{g}(\sigma')\,.
\end{equation}
It follows that $\bar{g}(\dsigma \cup \dsigma' \cup \sigma) \cup \bar{g}(\sigma) = \bar{g}(\dsigma \cup \sigma \cup \dsigma' \cup \sigma) = \bar{g}(\dsigma \cup \sigma)\cup\bar{g}(\dsigma' \cup \sigma)$,
so that $\fbar(\dsigma \cup \dsigma' \cup \sigma) + \fbar(\sigma) = \fbar(\dsigma \cup \sigma) + \fbar(\dsigma' \cup \sigma)$, 
and hence Eq.~\eqref{eqn:subadd} is satisfied at the boundary of the inequality, 
providing a relatively tight upper bound for Eq.~\eqref{eqn:boundobj}.

\section{Simulation framework}
\label{app:simfr}

The \texttt{Python3} simulation framework is structured into two main components: 
First, a ``\texttt{hepgk}'' module, implementing a small Constructive Solid Geometry (CSG) geometry kernel for representing the detector 
volume, and a simple vertex representation of tracking surfaces, as well as the core geometry optimizer classes which derive from the \texttt{pybnb}~\cite{pybnb} library.
Second, a ``\texttt{hepymc}'' module, that contains pythonic convenience wrapper derived from the \texttt{HepMC3}~\cite{hepmc3} python bindings,
and so-called processor classes, designed to reweight LLP events, implement particular LLP decay morphologies, 
and compute track hits on geometric elements, as required.
In this Appendix we summarize the general features, functionalities and approximate code flow of this framework.

\subsection{Geometry kernel}

Using the \texttt{hepgk} module, one can initialize different detector element geometries, 
including arbitrary fiducial decay volumes and tracking stations. 
Volumes are implemented at a fundamental level as unions, intersections or complements of convex polytopes 
that are fully specified by a list of their vertices. Additional sets of vertices encoding 2-dimensional polygons embedded in
three dimensions are used to specify tracking station panel elements. 

Several predefined geometries are implemented, 
that admit definitions in terms of the center, symmetry axes, and pertinent dimensions. 
For instance, the inputs for the (approximate polyhedral) \textlst{Cylinder} object are its center, axis, radius, and length. 
Other predefined polyhedra include \textlst{Cube}, \textlst{Tetrahedron} and \textlst{RectangularPrism}.
The user may also easily define additional preset geometries.

As an example, the nominal \CODEXb volume is implemented as
\begin{lstlisting}
import hepgk as gk

CODEXb_center, CODEXb_sidelen = [31e3, -2e3, 10e3], 10e3
CODEXb_geometry = gk.Cube(CODEXb_center, CODEXb_sidelen)
\end{lstlisting}

\subsection{HepyMC module and processors}

The \texttt{hepymc} module is built around the general \textlst{Processor} class. This abstract class takes lists of events and 
then processes or modifies them in some way, returning the processed list to an output HepMC file. 
More precisely, the processor returns an iterable generator which iterates over the events of the processed list. 
The user can create a specific processor by initializing a class that inherits from the parent \textlst{Processor} class. 
The currently available set of processor classes, their corresponding functionalities, and schematic inputs and outputs are shown in Table~\ref{tab:procs}.

\begin{table}
\renewcommand*{\arraystretch}{1.1}
\newcolumntype{C}{ >{\centering\arraybackslash\fontsize{10}{8}\selectfont } m{3.25cm} <{}}
\newcolumntype{D}{ >{\centering\arraybackslash\ttfamily} m{3cm} <{}}
\newcolumntype{E}{ >{\raggedright\arraybackslash\fontsize{10}{8}\selectfont } m{4.25cm} <{}}
\begin{tabular*}{0.9\linewidth}{@{\extracolsep{\fill}}DECC}
	\hline
	\textrm{Processor} & Functionality & Input & Output \\
	\hline\hline
	UnitSetter & sets energy and length units & HepMC event & HepMC event\\
	Splitter & partitions events generated (for convenience) with multiple LLP-producing decays into separate events with a single LLP-producing decay each & HepMC event & HepMC event(s) \\
	PhiStatBoost & increases MC statistics via discrete azimuthal rotation of events into the detector acceptance & HepMC event + detector geometry & HepMC event(s) \\
	Flipper & doubles MC statistics via discrete reflection in plane defined by the beam axis & HepMC event + detector geometry & HepMC event(s) \\ 
	Thrower & abstract \texttt{Processor} class for generation of weighted MC events, 
	subject to abstracted pre-~and post-checks & & \\
	Rescaler & \texttt{Thrower} class: generates LLP decay vertices (log uniform distributed) & HepMC event & HepMC event(s) + LLP decay vertex  \\
	VolumeFilter &  Selects events requiring the LLP decay vertex to be within a specified volume & HepMC event + LLP decay vertex + detector geometry & HepMC event + LLP decay vertex \\
	Decayer & \texttt{Thrower} class: generates 2-body LLP decays and attaches decay products to LLP decay vertex & HepMC event + LLP decay vertex & HepMC event(s) + LLP decay vertex + tracks\\
	Sampler & \texttt{Thrower} class: samples from existing set of simulated multibody decays and attaches decay products to LLP decay vertex & 
		HepMC event + LLP decay vertex + MC multibody decay sample & HepMC event(s) + LLP decay vertex + tracks \\
	PanelTracker & determines hits of tracks on tracking panel elements & HepMC event + LLP decay vertex + tracks + panel geometry & HepMC event + LLP decay vertex + tracks + hits  \\
	\hline	
\end{tabular*}
\renewcommand*{\arraystretch}{1.}
\caption{Available \textlst{Processor} classes, their functionalities, and schematic inputs and outputs.}
\label{tab:procs}
\end{table}

For example, the \textlst{VolumeFilter} processor selects all events from an \textlst{EventRecord} 
that contain an LLP that decays within a specified \texttt{hepgk} detector geometry.
It is implemented as
\begin{lstlisting}
from .processor import *

class VolumeFilter(Processor):
    def __init__(self, geometry, target=lambda p: p.pid == 999999,\
                precheck=lambda e: True, pstcheck=lambda e: True):

        self.geometry = geometry
        self.is_target = target
        super().__init__(self.identity, precheck=precheck,
                         pstcheck=lambda e: pstcheck(e) and self.in_vol(e))

    def identity(self, event):
        yield event

    def in_vol(self, event):
        return any(self.geometry.contains(p.end_vertex.position.vec3())\
                for p in event.particles if self.is_target(p)\
                    and p.end_vertex.position is not None)
\end{lstlisting}

To initialize the \textlst{VolumeFilter}, one must specify a \textlst{geometry} that determines the fiducial volume, 
as well as a \textlst{target} selection function for the decaying LLP particle of interest (by default assigned a particle identification (PID) number of \textlst{999999}). 
One can also pass additional so-called \textlst{precheck} or \textlst{pstcheck} condition checking functions, discussed briefly further below.
Initialization is completed by defining an \textlst{action} for the superclass:
In this case, it is the \textlst{identity} generator method, which simply yields an unmodified \textlst{event}.
The parent \textlst{Processor} class only performs \textlst{action} for events that pass the selections based on \textlst{precheck} and \textlst{pstcheck} functions,
which filter input and output events of the \textlst{action}, respectively.
Here, the default \textlst{pstcheck} uses the \textlst{in_vol} method which contains all the machinery of this processor, 
checking if an LLP decay vertex (whose location has typically has been set by a \textlst{Rescaler} preprocessor) is contained in the specified geometry.
Processor classes may also augment events or add events (that is, clone events from one to many) by implementing a less trivial \textlst{action} generator function. 
For example, the \textlst{Flipper} class applies a $Z_2$ reflection about a user-specified plane. 

\begin{lstlisting}
from .processor import *
from .. import event as ev, particle as pt
import numpy as np

class Flipper(Processor):
    def __init__self, axis='z', orientation=1, check_orientation=True,
                    precheck=lambda e: True, pstcheck=lambda e: True):
                    
        self.axis = {a : i for i, a in enumerate(['x', 'y', 'z', 't'])}[axis]
        self.orientation = np.sign(orientation)
        self._should_check = check_orientation
        self.PIDs = PIDs
        super().__init__(self.flip, precheck=precheck, 
                            pstcheck=lambda e: self.oriented(e) and pstcheck(e))

    def factor(self):
        return 2

    def oriented(self, event):
        llps = [p for p in event.particles if p.pid==999999]
        return any([self.orientation * llp.momentum[self.axis] >= 0 for llp in llps]) if self._should_check else True

    def flip(self, event):
        yield ev.Event(event)
        yield ev.Event(event).reflect(self.axis)
\end{lstlisting}
where the \textlst{reflect} method for events is provided by a wrapped \texttt{HepMC3} binding.

\subsection{Code flow}

In this framework, detector analysis proceeds via the following code flow:
\begin{enumerate}[label = \textbf{(\arabic*)}, noitemsep, topsep =0pt]
	\item An initial HepMC sample  is generated by \texttt{Pythia8}, (usually) containing undecayed LLPs, and stored.
	\item A detector geometry is initialized, specifying both a fiducial decay volume and tracking elements.
	\item These events are then passed through:
	\begin{enumerate}[label = \textbf{(\roman*)}, noitemsep, topsep =0pt]
		\item a units processor, to ensure correct conventions
		\item if needed, a splitter processor, to partition events with multiple LLP-producing decays into separate events 
		with a single LLP-producing decay each (an example is an event containing $B \bar{B}$ pairs, that are both decayed to an LLP by \texttt{Pythia8}).
		\item a rotation and reflection processor, to boost the statistical sample in the detector acceptance
		\item a reweighting processor, that assigns an event weight for the displacement of the LLP decay-in-flight, 
		according to an exponential with characteristic displacement $\beta\gamma c \tau$
		\item the fiducial volume processor, that selects events with LLP decays in the detector
		\item a decayer processor, that appends various desired final states to the LLP decay (such as $e^+e^-$ or $4\pi$ etc)
		\item a tracking processor, which appends track hit information on the various tracking elements to the HepMC record.
	\end{enumerate}
\end{enumerate}

\subsection{Normalization of efficiencies}
\label{app:norm}

For an MC sample of $N_s$ weighted events, indexed by $i$ with MC generation weight $w_i$, each containing a single LLP, 
the total efficiency (see definition at the beginning Sec~\ref{sec:reqs}) 
for the LLP decay vertex and kinematics to satisfy the acceptance and reconstruction requirements
has the simple estimator
\begin{equation}
	\varepsilon = \frac{1}{N_s} \sum_{i = 1}^{N_s} w_i r_{i,\text{dec}}\,.
\end{equation}
Here $r_{i,\text{dec}}$ is a reweighting factor for the LLP decay to fall in the detector acceptance (and the vertex be reconstructed): 
the ratio of the probability of decay (and reconstruction), $p_{i,\text{dec}}$, to the sampling probability, $p_s$, 
which is usually taken to be generated according to a log-uniform distribution in the LLP decay vertex displacement.
I.e., $r_{i,\text{dec}} = p_{i,\text{dec}}/p_s$.
The numerator is zero if the LLP does not decay in the detector acceptance (and/or is not reconstructed).
Often the LLPs are generated in unweighted events, i.e. $w_i = 1$; but in cases that, e.g., 
a long tail in the $p_T$ of the LLP parent must be sampled efficiently, 
the MC generation may assign a nontrivial weight.

Further, because LLPs may be multiply produced in some portals, such as in $h \to A'A'$, 
in computing the efficiency for these portals one must be careful with combinatorics and correlations.
In this scenario, we are interested in computing the efficiency for \emph{at least} one LLP to decay and be reconstructed by the detector in any given event.
For an MC sample of $N_s$ weighted events containing $n_i$ LLPs in the $i$th event, 
the $i$th (estimator of the) reweighting factor for detecting at least one LLP decay then becomes $r_{i,\text{dec}} = [1 - \prod_{\alpha=1}^{n_i}(1 - p_{\alpha,\text{rec}})]/p_s$.
Because typically $p_{\alpha,\text{rec}} \ll 1$ in the detectors we consider, 
then to a good approximation this may be linearized over the LLPs, so that
\begin{equation}
	\varepsilon \simeq  \frac{1}{N_s} \sum_{i =1}^{N_s} w_i \sum_{\alpha =1}^{n_i} r_{\alpha,\text{dec}}\,,
\end{equation}
with $r_{\alpha,\text{dec}} = p_{\alpha,\text{rec}}/p_s$.

For convenience and parallelization, our MC samples are partitioned into subsamples.
Thus to properly normalize the LLP event weights in order to compute the total efficiency, 
one must keep track of the total $N_s$ across multiple subsamples:
if $N_s$ is different in each subsample, then the average of the ratios is not the ratio of the sum.
Further, techniques that boost the MC statistics, such as discrete azimuthal rotation or reflection classes, 
modify the effective number of events in the simulation (sub)sample, 
and therefore require a careful accounting.
The framework addresses both of these effects by allowing creation of a special final event in each output HepMC (sub)sample, 
that keeps track of the number of discarded events, $N_{\text{disc}}$, that \emph{did not} pass selection by whatever processor was applied to the (sub)sample.
The sum of the $N_{\text{disc}}$ plus the total number of events in set of subsamples determines the denominator in the fiducial efficiency.

As an example, we may consider the normalization logic for the \textlst{PhiStatBoost} processor, 
that implements an enhancement of the MC statistics through discrete azimuthal angular rotations.
In the particular case of the \CODEXb detector, 
one observes that the detector acceptance subtends quite a small angle around the beamline, just less than $2\pi/16$ radians.
Thus, if one takes an initial MC sample and applies discrete azimuthal rotations in steps of $2\pi/16$, 
keeping all events with at least one LLP in the detector azimuthal angular acceptance at each step, 
one may significantly boost the statistics by approximately a factor of $16$.
The processor logic works as follows:
\begin{enumerate}[label = \textbf{(\arabic*)}, noitemsep, topsep =0pt]
	\item Compute the minimal azimuthal angular wedge of size $2\pi/n$, i.e. for maximal possible integer $n$, covering the detector acceptance (for \CODEXb, $n = 16$).\\
	Then, for each event:
	\item Extract the LLP momenta (there may be more than one, in the case of e.g. $h \to A'A'$).
	\item \label{it:chkang} If at least one LLP momentum lies in the detector angular acceptance, write the event to the output sample. 
	Otherwise increment the discarded events $N_{\text{disc}}$ by one.
	\item Rotate the entire event by $2\pi/16$ and return to step~\ref{it:chkang}, until $n$ rotations have been performed. 
\end{enumerate}
Examining a particular sample of $50000$ $h \to A'A'$ events as an example,
we naively expect only $50000/16\times2 = 6250$ events to contain at least one LLP inside the detector acceptance.
Applying the \textlst{PhiStatBoost} processor, one finds it is processed into a sample of $N_s = 99343$ events
with at least one LLP inside the detector acceptance, and $N_{\text{disc}} = 700657$; the total is $16 \times 50000$. 
Thus we see the statistics is boosted by approximately a factor of $99343/6250 \simeq 15.9$ as expected.
Combining $M$ such rotated subsamples, the total efficiency estimator becomes
\begin{equation}
	\varepsilon = \Big(\sum_{m =1}^M \sum_{i = 1}^{N_{m,s}} w_{m,i} r_{m,i,\text{dec}}\Big)\Big/\Big[\sum_{m =1}^M N_{m,s} + N_{m,\text{disc}}  \Big]\,.
\end{equation}

\subsection{Example scripts}

Here we show an example script that implements the \textlst{PhiStatBoost} processor, 
taking a HepMC subsample file (or list of files) and geometry as a command line argument,
and writing the final normalization event in the output HepMC.

\begin{lstlisting}
mport os, sys, argparse as ap, pickle as pk, itertools as it
from glob import glob
from tqdm import tqdm
sys.path.append('./')
sys.path.append('../')
import hepymc as mc, hepymc.processors as proc

from pyHepMC3 import HepMC3 as hmc3
hmc3.Setup.set_print_warnings(False)

if __name__ == "__main__":
    parser = ap.ArgumentParser(description="Provide statistics boost by rotational symmetry in `.hepmc` files")
    parser.add_argument("geometry", metavar="GEO_FILE",
                        help="Pickled `.pk` geometry file to base phi angle on.",
                        type=str)
    parser.add_argument("files", metavar="EVENT_FILE",
                        help="Event `.hepmc` file(s) to process.",
                        nargs='+', type=str, default=[])
    parser.add_argument("-v", "--verbose", help="Indicate status.",
                        action="store_true")
    parser.add_argument("-l", "--logged", help="Indicate for logged use.",
                        action="store_true")
    args = parser.parse_args()

    with open(args.geometry, 'rb') as f:
        geometry, _ = pk.load(f)
    rotbst = proc.PhiStatBoost(geometry, check_orientation=True)

    filenames = [name for fn in args.files for name in glob(fn)]
    for fn in tqdm(filenames, desc="File", unit="files",
                        leave=args.logged, disable=not args.verbose):

        with mc.open(fn, 'r', version=3) as f:
            events = list(tqdm(f, desc="Loading", unit="events", leave=args.logged,
                            disable=not args.verbose, postfix=os.path.basename(fn)))

        events = list(rotbst(tqdm(events, desc="Boosting", unit="events", leave=args.logged,
                            disable=not args.verbose, postfix="x" + str(rotbst.factor))))

        fo = os.path.splitext(fn)[0] + ".ROTBSTx" + str(rotbst.factor) + ".hepmc3"
        with mc.open(fo, 'w', version=3) as f:
            f.write_events(tqdm(events, desc="Writing", unit="events",
                                leave=args.logged, disable=not args.verbose,
                                postfix=os.path.basename(fo)))
            pre_discarded_count = proc.bin_events(tqdm(rotbst.prediscarded,
                                        desc="Accounting pre-discarded",
                                        leave=args.logged, disable=not args.verbose),
                                        factor=lambda e: rotbst.factor)
            discarded_count = proc.bin_events(it.chain(pre_discarded_count,
                                        tqdm(rotbst.pstdiscarded,
                                        desc="Accounting post-discarded", unit="events",
                                        leave=args.logged, disable=not args.verbose)))
            f.write_events(tqdm(discarded_count,
                            desc="Writing Normalization", unit="events",
                            leave=args.logged, disable=not args.verbose))
        rotbst.reset()
        del events
\end{lstlisting}

In addition, we show a script that implements the \textlst{Rescaler} processor.
This processor computes the reweighting factors for LLP decays with respect to a log uniform sampling distribution in the LLP displacement $l = 10^u c\tau$,
for a range of $\log(c\tau/\text{mm})$ values. 
These are passed as a list to the command line \textlst{"--ctaus"} argument.
The range of the log uniform sampling distribution exponent $u$ is defined by arguments \textlst{"--minlog"} and \textlst{"--maxlog"},
The number of sampling ``throws'' for each LLP is determined by the argument  \textlst{"--num"}; a clone of the event is created for each throw
and written to output HepMC.

\begin{lstlisting}
#!/usr/bin/python3
import os, sys, argparse as ap, itertools as it
from glob import glob
from tqdm import tqdm
sys.path.append('./')
sys.path.append('../')
import hepymc as mc, hepymc.processors as proc

from pyHepMC3 import HepMC3 as hmc3
hmc3.Setup.set_print_warnings(False)

if __name__ == "__main__":
    parser = ap.ArgumentParser(description="Rescale target particle lifetimes in `.hepmc` files")
    parser.add_argument("files", metavar="EVENT_FILE",
                        help="Event `.hepmc` file(s) to process.",
                        nargs='+', type=str, default=[])
    parser.add_argument("-t", "--ctaus", help="Log lifetimes (in event current units) to scale to.", 
                        nargs='+', type=float, default=[])
    parser.add_argument("--linthrow", help="Interpret log ctaus as linear lifetimes in the event current units.",
                        action="store_false")
    parser.add_argument("-n", "--num", help="Number of repeated throws to make per event.",
                        type=int, default=1)
    parser.add_argument("-p","--pids", help="PDG PIDs to target, default=[999999].",
                        nargs='+', type=int, default=[999999])
    parser.add_argument("--minlog", help="Min log throw.", type=float, default=-3.)
    parser.add_argument("--maxlog", help="Max log throw.", type=float, default=11.)
    parser.add_argument("--noprop", help="Don't propagate rescaling to children.", 
                        action="store_true")
    parser.add_argument("-v", "--verbose", help="Indicate status.",
                        action="store_true")
    parser.add_argument("-l", "--logged", help="Indicate for logged use.",
                        action="store_true")
    args = parser.parse_args()

    rescale = proc.Rescaler(args.ctaus, logt=args.linthrow, PID=args.pids, num=args.num,
        minlogthrow=args.minlog, maxlogthrow=args.maxlog, prop=not args.noprop)

    filenames = [name for fn in args.files for name in glob(fn)]
    for fn in tqdm(filenames, desc="File", unit="files",
                        leave=args.logged, disable=not args.verbose):

        with mc.open(fn, 'r', version=3) as f:
            events = list(tqdm(f, desc="Loading", unit="events", leave=args.logged,
                            disable=not args.verbose, postfix=os.path.basename(fn)))

        events = list(rescale(tqdm(events, desc="Rescaling", unit="events", leave=args.logged,
                            disable=not args.verbose)))

        fo = os.path.splitext(fn)[0] + ".RESCALED" + str(args.ctaus).strip() + ".hepmc3"
        with mc.open(fo, 'w', version=3) as f:
            f.write_events(tqdm(events, desc="Writing", unit="events",
                                leave=args.logged, disable=not args.verbose,
                                postfix=os.path.basename(fo)))
            pre_discarded_count = proc.bin_events(tqdm(rescale.prediscarded,
                                        desc="Accounting pre-discarded",
                                        leave=args.logged, disable=not args.verbose))
            discarded_count = proc.bin_events(it.chain(pre_discarded_count,
                                        tqdm(rescale.pstdiscarded,
                                        desc="Accounting post-discarded", unit="events",
                                        leave=args.logged, disable=not args.verbose)))
            f.write_events(tqdm(discarded_count,
                            desc="Writing Normalization", unit="events",
                            leave=args.logged, disable=not args.verbose))
        rescale.reset()
        del events
\end{lstlisting}

\bibliographystyle{JHEP}

\providecommand{\href}[2]{#2}\begingroup\raggedright\endgroup

\end{document}